\documentclass[10pt,halfline,a4paper]{ouparticle}

\usepackage{graphicx}
\usepackage{multirow}
\usepackage{subfigure}
\usepackage{cite}
\usepackage{algorithm}
\usepackage{algorithmic}
\usepackage{amsmath}
\usepackage{amsfonts}
\usepackage{amssymb}
\usepackage{booktabs, color}
\usepackage{amsmath,amssymb,amsfonts}
\usepackage{mathrsfs}

\usepackage[aboveskip=1pt,labelfont=bf,labelsep=period,justification=raggedright,singlelinecheck=off]{caption}

\usepackage{pifont}% http://ctan.org/pkg/pifont

\newcommand{\real}{\mathbb{R}}

\newcommand{\transpose}{\mathsf{T}} %or \transpose or \intercal
\newcommand{\mc}{\mathcal}

\newcommand{\until}[1]{\{1,\dots,#1\}}
\newcommand{\fromto}[2]{\{#1,\dots,#2\}}

\newcommand{\1}{\boldsymbol{1} }
\DeclareMathOperator{\diag}{diag}

\newcommand*{\QEDA}{\hfill\ensuremath{ acksquare}}%
\newenvironment{pfof}[1]{\vspace{1ex}\noindent{\itshape Proof of
    #1:}\hspace{0.5em}} {\hfill\QEDA\vspace{1ex}}

\newtheorem{appxthm}{Theorem}[section]

\newcommand{\rd}{\color{red}}

\newcommand{\bs}{\boldsymbol}

\usepackage[left=.75in,right=.75in,top=1in,bottom=1in]{geometry}

\begin{document}

\title{Functional Control of Oscillator Networks}

\author{%
\name{Tommaso Menara}
\address{Department of Mechanical and Aerospace Engineering, University of California, San Diego, La Jolla, CA 92093, USA}
\and
\name{Giacomo Baggio}
\address{Department of Information Engineering, University of Padova, Padova, 35131, Italy}
\and
\name{Danielle S. Bassett}
\address{Department of Physics \& Astronomy,
Department of Bioengineering,
Department of Electrical \& Systems Engineering,
Department of Neurology,
Department of Psychiatry, University of Pennsylvania, Philadelphia, PA 19104, USA\\
The Santa Fe Institute, Santa Fe, NM 87506, USA}
\and
\name{Fabio Pasqualetti}
\address{Department of Mechanical Engineering, University of California, Riverside, Riverside, CA 92521, USA}
\email{fabiopas@engr.ucr.edu}}

\date{}

\maketitle
\newpage
\section*{Abstract}
Oscillatory activity is ubiquitous in natural and engineered network systems. { The interaction scheme underlying interdependent oscillatory components governs the emergence of network-wide patterns of synchrony that regulate and enable complex functions. Yet, understanding, and ultimately harnessing, the structure-function relationship in oscillator networks {remains an outstanding challenge} of modern science. Here, we address this challenge by presenting a principled method to prescribe exact and robust functional configurations from local network interactions through optimal tuning of the oscillators' parameters.}  To quantify the behavioral synchrony between { coupled oscillators,} we introduce the notion of \emph{functional pattern}, which encodes the pairwise relationships between the oscillators' phases.  { Our procedure is computationally efficient and provably correct, accounts for constrained interaction types, and allows to concurrently assign multiple desired functional patterns. Further, we derive algebraic and graph-theoretic conditions to guarantee the feasibility and stability of target functional patterns. These conditions provide an interpretable mapping between the structural constraints and their functional implications in oscillator networks.}
As a proof of concept, we apply the proposed method to replicate empirically recorded functional relationships from cortical oscillations in a human brain, and to redistribute the active power flow in { different models of electrical grids}.

\section*{Introduction}

Complex coordinated behavior of oscillatory { components is linked
  to the function} of many natural and technological network systems \cite{MCC-PCH:93,AP-MR-JK:03,AC-AC-IG-GP-FS-MV:10}. For instance, {
  distinctive network-wide patterns of synchrony
  \cite{IS-MG-MP:03,MG-IS-AT:05,FS-EO:07}} determine the coordinated motion of orbiting particle systems \cite{FZ-NEL:07}, promote successful mating in populations of fireflies \cite{JB-EB:66}, regulate the active power flow in electrical grids \cite{FD-MC-FB:13}, predict global climate change phenomena \cite{KET:76}, dictate the structural {
  development} of mother-of-pearl in mollusks
\cite{MB-DZ-AP-PZ-IZ:21}, and enable numerous cognitive functions in
the brain \cite{JC-EH-OS-GD:11,FJ-AN:2011}. { Since this rich
  repertoire of patterns emerges from the properties of the underlying
  interaction network \cite{SHS:01}, controlling the collective {
    configuration} of interdependent units holds a tremendous
  potential across science and engineering \cite{SHS-IS:93}.
  Despite its practical significance, a comprehensive method to
  enforce network-wide patterns of synchrony by intervening on the
  network's structural parameters does not yet exist.

  In this work, we develop a rigorous framework that allows us to optimally control
  the spatial organization of the network components and their
  oscillation frequencies to achieve desired patterns of synchrony. We abstract the
  rhythmic activity of a system as the output of a network of
  diffusively coupled oscillators
  \cite{AA-ADG-JK-YM-CZ:08,Doerfler2014} with Kuramoto dynamics. This modeling choice is motivated by the rich dynamical repertoire and wide adoption of Kuramoto oscillators \cite{JAA-LLB-CJPV-FR-RS:05}. Specifically, we consider an undirected network $\mc G = \{\mc O, \mc E\}$ of $n$ oscillators with dynamics
 \begin{align}\label{eq: kuramoto}
  \dot \theta_i = \omega_i + \sum_{j=1}^n A_{ij} \sin(\theta_{j}-\theta_{i}),
\end{align}
where $\omega_i\in\real$ and $\theta_i\in\mathbb{S}^1$ are the
frequency and phase of the $i$-th oscillator, respectively,
$A = [A_{ij}]$ is the weighted adjacency matrix of $\mc G$, and
$\mc O = \until{n}$ and $\mc E\subseteq\mc O\times \mc O$ { denote}
the oscillator and { interconnection} sets, respectively. In this
work we consider the case where the network $\mc G$ admits both
cooperative (i.e., $A_{ij}>0$) and competitive (i.e., $A_{ij}<0$)
\cite{HH-SHS:11} interactions among the oscillators, as well as the
more constrained case of purely cooperative interactions that arises
in several real-word systems. For instance, negative interactions are
not physically meaningful in networks of excitatory
neurons, in power distribution networks (where the
  interconnection weight denotes conductance and susceptance of a
transmission line), and in urban transportation networks (where
interconnections denote the number of vehicles on a road with respect to its maximum capacity).

To quantify the pairwise functional relations between oscillatory units, and inspired by the work in Ref.~\cite{AA-AD-CJP:06}, we define a local correlation metric { that, compared to the classical Pearson correlation coefficient, does not depend on sampling time and is more convenient when dealing with periodic phase signals (see Supplementary Information)}.
 Given a pair of phase oscillators $i$ and $j$ with phase trajectories $\theta_i(t)$ and $\theta_j(t)$, we define the correlation coefficient
\begin{equation}\label{eq: corr coeff}
\rho_{ij} = <\cos(\theta_j(t)-\theta_i(t))>_t,
\end{equation}
where $<\cdot>_t$ denotes the average over time. A \emph{functional
  pattern} is formally defined as the symmetric matrix $R$ whose
$i,j$-th entry equals $\rho_{ij}$. Importantly, a functional pattern
explicitly encodes { the pairwise, local, correlations across all
  of the} oscillators, which { are more informative than} a global
observable (e.g., the order parameter
\cite{AA-ADG-JK-YM-CZ:08,LA-AD-AA:18}). It is easy to see that, if two
oscillators $i$ and $j$ synchronize after a certain initial transient,
$\rho_{ij}\to 1$ as time increases. If two oscillators $i$ and $j$
become phase-locked (i.e., their phase difference remains constant
over time), { then} their correlation coefficient converges to some
constant value { with magnitude smaller than 1}. If the phases of
two oscillators $i$ and $j$ evolve independently, { then} their
correlation value { remains} small over time. A few questions arise
naturally, which will be answered in this paper. Are all functional
patterns achievable? Which network configurations allow for the
emergence of { multiple} target functional patterns? And, if a
certain functional pattern can be achieved, is it robust to
perturbations? Surprisingly, we reveal that controlling functional
patterns { can be cast as} a convex optimization problem, whose
solution can be characterized explicitly. { Fig.~\ref{fig:
    conceptual} shows our framework and an example of control of
  functional pattern for a network with $7$ oscillators.} In the
paper, we will validate our
  methods by replicating functional patterns from brain recordings in
  an empirically reconstructed neuronal network, and by controlling
  the active power distribution in multiple models of power grid.

\begin{figure}[t]
\centering
 \includegraphics[width=1\columnwidth]{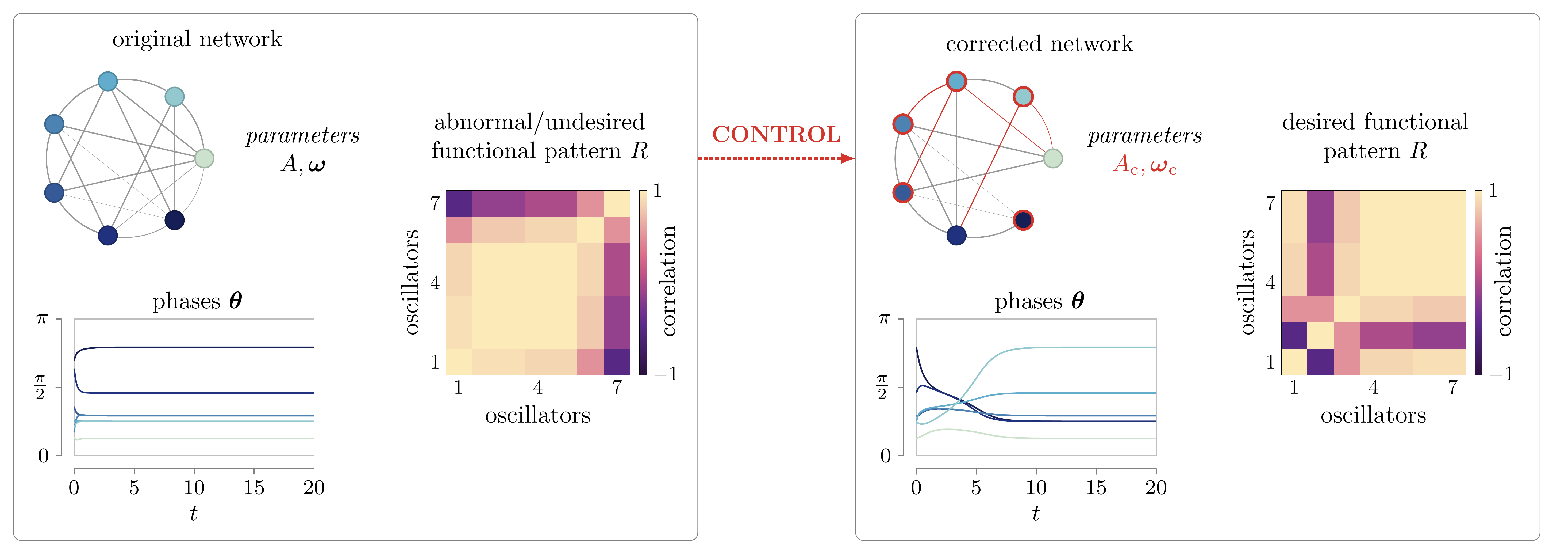}%
  \vspace{0.2cm}
 \caption{\textbf{Network control to enforce a desired functional pattern from an abnormal or undesired one.} The left panel contains a network of $n=7$ oscillators (top left panel, line thickness is proportional to the coupling strength), whose vector of natural frequencies ${\bs\omega}$ has zero mean. The phase differences with respect to $\theta_1$ (i.e., $\theta_i-\theta_1$) converge to $\left[\frac{\pi}{8}~\frac{\pi}{8}~ \frac{\pi}{6}~ \frac{\pi}{6}~ \frac{\pi}{3}~ \frac{2\pi}{3}\right]$, as also illustrated in the phases' evolution from random initial conditions (center left panel, color coded). The bottom left panel depicts the functional pattern $R$ corresponding to such phase differences over time. The right panel illustrates the same oscillator network after a selection of coupling strengths and natural frequencies have been tuned (in red, the structural parameters $A$ and $\bs \omega$ are adjusted to { $A_\mathrm{c}$ and ${\bs \omega}_\mathrm{c}$}) to obtain the  phase differences $\left[\frac{2\pi}{3}~ \frac{\pi}{3}~ \frac{\pi}{6}~ \frac{\pi}{6}~ \frac{\pi}{8}~ \frac{\pi}{8}\right]$, which encode the desired functional pattern in the bottom right. In this example, we have { computed} the closest set (in the $\ell_1$-norm sense) of coupling strengths and natural frequencies to the original ones that enforce the emergence of the { target} pattern. Importantly, only a subset of the original parameters has been modified. \label{fig: conceptual}}
 \end{figure}

 While synchronization phenomena in oscillator networks have been
 studied extensively (e.g., see {
   \cite{Pecora1998,SHS:03,FS-MDB-FG:07,NT-AEM:10,IB-RJ-VB:17}}), the
 development of control methods to impose desired synchronous
 behaviors has only recently attracted the attention of the research
 community \cite{LP-JZK-JK-DSB:17,MF-FD-VMP:17, AF-FGW-JD:18,
   TM-GB-DSB-FP:19b}. Perhaps the work that is closest to our approach
 is Ref.~\cite{MF-FD-VMP:17}, where the authors tailor interconnection
 weights and natural frequencies to achieve a specified level of phase
 cohesiveness in a network of Kuramoto oscillators. Our work improves
 considerably upon this latter study, whose goal is limited to
 prescribing an upper bound to the phase differences, by enabling the
 prescription of pairwise phase differences and by investigating the
 stability properties of different functional patterns. Taken
 together, existing results highlight the importance of controlling
 distinct configurations of synchrony, but remain mainly focused on
 the control of ``macroscopic'' observables of synchrony (e.g., the
 average synchronization level of all the oscillators). In contrast,
 our control method prescribes desired pairwise levels of correlation
 across all of the oscillators, thus enabling a precise ``microscopic''
 description of functional interactions.}

\section*{Analytical results}

\subsection*{Feasible { functional patterns} in positive networks}

{ A functional pattern is an $n\times n$ matrix whose entries are the
time-averaged cosine of the differences of the oscillator phases (see
equation~\eqref{eq: corr coeff}). When the oscillators reach an
equilibrium, the differences of the oscillator phases become constant, and
the network evolves in a \emph{phase-locked} configuration. In this
case, the functional pattern of the network also becomes constant,
and is uniquely characterized by the phase differences at the
equilibrium configuration. In this work we study functional patterns
for the special case of phase-locked oscillators and, given the
equivalence between a desired functional pattern and a desired set of
phase differences at equilibrium, convert the problem of generating a
functional pattern into the problem of ensuring a desired phase-locked
configuration. We recall that, while convenient for the analysis,
phase-locked configurations play a crucial role in the functioning of
many natural and man-made networks
\cite{SK-MRP-WHR-JTS-SEK-JAK:05,MvW-MV-JL-CMAP:10,CK-MT-DB:16}.

For the undirected network $\mc G = (\mc O, \mc E)$, let
$x_{ij} = \theta_j - \theta_i$ be the difference of the phases of the
oscillators $i$ and $j$, and let $\bs x\in \real^{|\mc E|}$ be the
vector of all phase differences with $(i,j) \in \mc E$ and $i<j$. The
network dynamics \eqref{eq: kuramoto} can be conveniently rewritten in
vector form as (see Methods)
\begin{align}\label{eq: kuramoto matrix}
  B D({\bs x}) \bs  \delta = [\omega_1\;~\cdots
  \;~\omega_n]^\transpose - \dot{ \bs  \theta},
\end{align}
where $B\in\real^{n\times |\mc E|}$ is the (oriented) incidence matrix
of the network $\mc G$, $D({\bs x})\in\real^{|\mc E|\times |\mc E|}$
is a diagonal matrix of the sine functions in equation~\eqref{eq:
  kuramoto}, and $\bs \delta\in\real^{|\mc E|}$ is a vector collecting
all the weights $A_{ij}$ with $i<j$. Because we focus on phase-locked
trajectories, all oscillators evolve with the same frequency and the
vector $\dot{\bs \theta}$ satisfies
$\dot{\bs \theta} = \omega_\text{mean} \textbf{1}$, where
$\omega_\text{mean} = \left(\frac{1}{n}\sum_{i=1}^n \omega_i\right)$
is the average of the natural frequencies of the oscillators. Further,
since $\mc G$ contains only $n$ oscillators, any phase difference
$x_{ij}$ can always be written as a function of $n-1$ independent
differences; for instance, $\{x_{12}, x_{23},\dots,x_{n-1,n}\}$. In
fact, for any pair of oscillators $i$ and $j$ with $i<j$, it holds
$x_{ij} = \sum_{k = i}^{j-1}x_{k,k+1}$. This implies that the vector
of all phase differences in equation~\eqref{eq: kuramoto matrix}, and
in fact any $n\times n$ functional pattern, has only $n-1$ degrees of
freedom and can be uniquely specified with a set of $n-1$ independent
differences $\bs x_\text{desired}$ (see Methods). Following this
reasoning and to avoid cluttered notation, let
$\bs \omega = [\omega_1-\omega_\text{mean} \; ~\cdots \;~\omega_n-
\omega_\text{mean}]^\transpose$, and notice that the problem of
enforcing a desired functional pattern simplifies to (i) converting
the desired functional pattern to the corresponding phase differences
$\bs x_\text{desired}$, and (ii) computing the network weights $\bs \delta$ to
satisfy the following equation:
\begin{equation}\label{eq: linear}
  B D({\bs x}) \bs \delta = \bs \omega,
\end{equation}
where we note that the vector $\bs\omega$ has zero mean and that, with
a slight abuse of notation, $D({\bs x})$ denotes the
$|\mc E|$-dimensional diagonal matrix of the sine of the phase
differences uniquely defined by the ($n-1$)-dimensional vector $\bs x_\text{desired}$.

We begin by studying the problem of attaining a desired functional
pattern using only nonnegative weights. With the above notation, for a
desired functional pattern corresponding to the phase differences
$\bs x$, this problem reads as}
\begin{align}\label{eq: problem 2}
  \mathrm{find }&&&   \bs  \delta \\
  \mathrm{subject~to }&&& B D(\bs x) \bs \delta = \bs \omega, \tag{\theequation a}\label{eq: constraint 1} \\
\text{ and }&&& \bs \delta \ge 0. \tag{\theequation b}\label{eq: positive constraint 1}
\end{align}
{ It should be noticed that the feasibility of the optimization problem \eqref{eq:
  problem 2} depends on the sign of the entries of the diagonal matrix
$D(\bs x)$, but is independent of their
magnitude. To see this, notice that
\begin{align*}
  D(\bs x) = \text{sign} (D(\bs
  x) ) | D( \bs
  x) |,
\end{align*}
where the $\mathrm{sign}(\cdot)$ and absolute value $|\cdot|$
operators are applied element-wise. Then, Problem \eqref{eq: problem
  2} is feasible if and only if there exists a nonnegative solution to
\begin{align*}
  \underbrace{B \text{sign} (D(\bs
  x) )}_{\bar B} \underbrace{| D( \bs
  x) | \bs \delta}_{\bar{\bs \delta}} = \bs \omega .
\end{align*}
The feasibility of the latter equation, in turn, depends on the
projections of the natural frequencies $\bs \omega$ on the columns of
$\bar B$: a nonnegative solution exists if $\bs\omega$ belongs to the
cone generated by the columns of $\bar B$. This also implies that, if
a network admits a desired functional pattern
$\bs x$ then, by tuning its weights, the same network
can generate any other functional pattern
${\bs x}_\text{new}$ such that
$\text{sign} (D({\bs x}_\text{new}) ) = \text{sign} (D({\bs
  x}) )$. Thus, by properly tuning its weights, a
network can generally generate a continuum of functional patterns
determined uniquely by signs of its incidence matrix and the oscillators natural
frequencies. This property is illustrated in Fig.~\ref{fig: mapping
  weights equilibria} for the case of a line network.

\begin{figure}[t]
\centering
 \includegraphics[width=1\columnwidth]{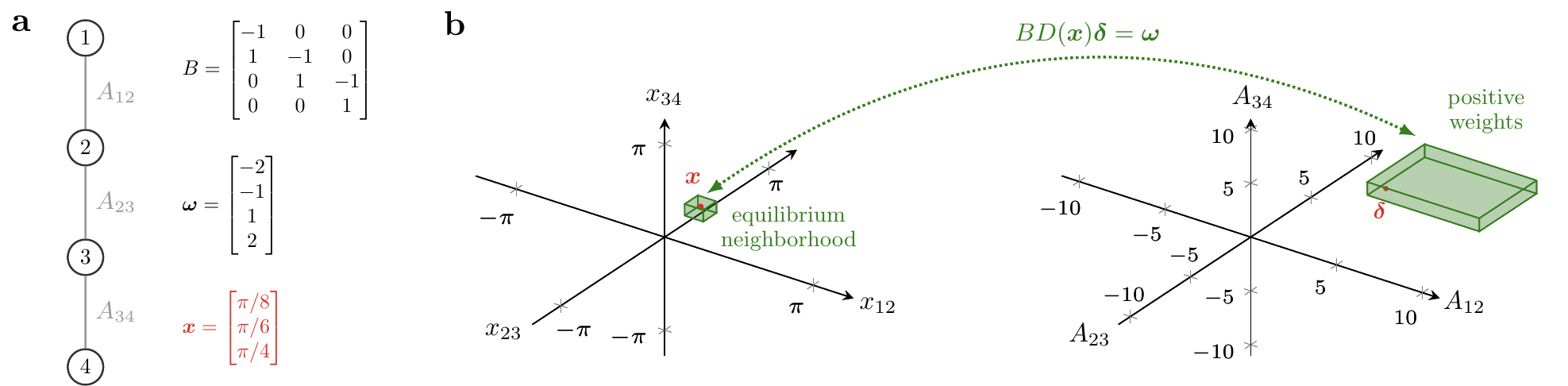}%
  \vspace{0.4cm}
 \caption{ \textbf{Mapping between desired phase differences and interconnection weights.} \textbf{a} A line network of $n=4$ nodes and its parameters. The desired phase differences are shown in red. \textbf{b} Left panel: space of the phase differences; right panel: space of the interconnection weights. The pattern ${\bs x}$ is illustrated in red in the left panel, and the network weights that achieve such a pattern are represented in red in the right panel. For fixed natural frequencies $\bs \omega$, the green parallelepiped on the left represents the continuum of functional patterns within $0.2$ radians from ${\bs x}$ which can be generated by the positive interconnection weights in the green parallelepiped on the right. \label{fig: mapping weights equilibria}}
 \end{figure}

A sufficient condition for the feasibility of Problem \eqref{eq:
  problem 2} is as follows: \medskip

\emph{There exists $\bs \delta \ge 0$ such that
  $ B D(\bs x)\bs \delta = { \bs \omega}$ if there
  exists a set $\mc S$ satisfying:} \begin{enumerate}
  \item[(i.a)]
  \emph{$D_{ii}( \bs
  x) D_{jj}( \bs
  x) B_{:,i}^\transpose B_{:,j} \le 0$ for all
    $i,j\in \mc S$ with $i\ne j$ and $D_{ii},D_{jj}\ne 0$;} 
    \item[(i.b)]
  \emph{${ \bs \omega}^\transpose B_{:,i}D_{ii}  (\bs x)> 0$ for all $i\in \mc S$;}
\item[(i.c)]
  \emph{${\bs \omega} \in \mathrm{Im}(B_{:,\mc S})$.}
\end{enumerate}

Equivalently, the above conditions ensure that ${ \bs \omega}$ is contained within the cone
generated by the columns of $\bar B_{:,\mc S}$ (see Fig.~\ref{fig: existence}\textbf{a} for a self-contained example). 
To see this, rewrite the pattern assignment problem $B D({\bs x}) \bs \delta = { \bs \omega}$ as 
\begin{align}\label{eq:solution-S}
  B D({\bs x}) \bs \delta = B_{:,\mc S}
  D_{\mc S,\mc S}({\bs x})\bs  \delta_{\mc S}  +
  B_{:,\tilde{\mc S}} D_{\tilde{\mc S},\tilde{\mc S}}({\bs x})\bs
  \delta_{\tilde{\mc S}} = \bs \omega,
\end{align}
where the subscripts $\mc S$ and $\tilde{\mc S}$ denote the entries
corresponding to the set $\mc S$ and the remaining ones,
respectively. If the vectors $B_{:,i}$,
  $i\in{\mc S}$, are linearly independent, condition (i.a) implies that
$D_{\mc S,\mc S} B_{:,\mc S}^{\transpose} B_{:,{\mc S}}D_{\mc S,\mc
  S}$ is an M-matrix; that is, a matrix which has nonpositive
off-diagonal elements and positive principal minors
\cite{RAH-CRJ:85}. Otherwise,
  the argument holds verbatim by replacing $\mc S$ with any subset
  $\mc S_{\text{ind}}\subset \mc S$ such that the vectors $B_{:,i}$,
  $i\in{\mc S_{\text{ind}}}$, are linearly independent. Condition
(i.c) guarantees the existence of a solution to
$B_{:,\mc S} D_{\mc S,\mc S} \bs\delta_{\mc S}=\bs\omega$. A
particular solution to the latter equation is
$$
  \bs  \delta_{\mc S} 
  = \left(B_{:,\mc S} D_{\mc S,\mc S}({\bs x})\right)^\dagger \bs \omega  = ( D_{\mc S,\mc S}({\bs x}) B_{:,\mc S}^{\transpose} B_{:,\mc S} D_{\mc S,\mc S}({\bs x}))^{-1} D_{\mc S,\mc S}({\bs x}) B_{\mc S,\mc S}^\transpose \bs \omega >0
$$
where $(\cdot)^\dagger$ denotes the Moore-Penrose pseudo-inverse of
a matrix. The positivity of $\bs  \delta_{\mc S}$ follows from condition (i.b) and the fact that the inverse of an M-matrix is element-wise nonnegative \cite{RAH-CRJ:85}. We conclude that a solution to equation~\eqref{eq:solution-S} is given by $\bs \delta_{\mc S}= \left(B_{:,\mc S} D_{\mc S,\mc S}({\bs x})\right)^\dagger \bs \omega>0$ and $\bs \delta_{\tilde{\mc S}}= \bs 0$. 

 \begin{figure}[t]
\centering
 \includegraphics[width=1\columnwidth]{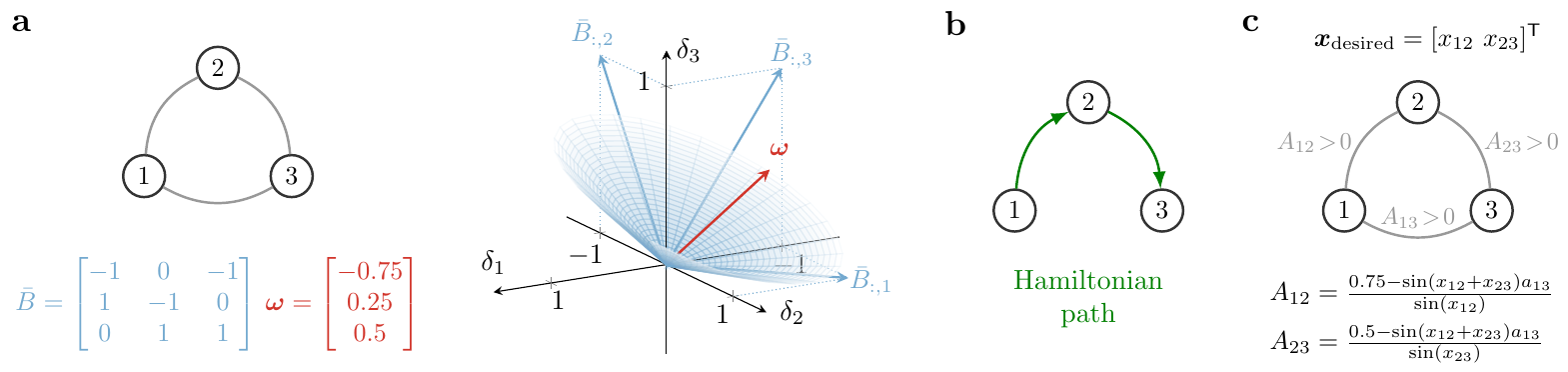}%
  \vspace{0.3cm}
 \caption{\textbf{Algebraic and graph-theoretic conditions for the existence of { positive weights that attain a desired functional pattern}.} \textbf{a} The left side illustrates a simple network of $3$ oscillators with adjacency matrix $\bar B$ and vector of natural frequencies ${ \bs \omega}$. The right side illustrates the cone generated by the columns of $\bar B$. In this example, $\mc S = \{1,2\}$ satisfies the conditions for the existence of $\bs \delta\ge 0$ in equation~\eqref{eq: problem 2}, as ${ \bs  \omega}$ is contained within the cone generated by the columns $\bar B_{:,\mc S}$. \textbf{b}  The (directed) Hamiltonian path described by the columns of $\bar B_{:,\mc H}$, with $\mc H = \{1,2\}$, in the network of panel \textbf{a}. \textbf{c} The existence of such an Hamiltonian path, together with a positive projection of ${\bs\omega}$ onto $\bar B_{:,\mc H}$, also ensure that there exists a strictly positive $\bs \delta > 0$ solution to $ B D({\bs x})\bs \delta = { \bs  \omega}$. In particular, for any choice of $ x_{12},  x_{23}\in{ (0,~\pi)}$, equation \eqref{eq: linear} reveals that if $0<A_{13}< 0.5/\sin( x_{12}+ x_{23})$, then there exist strictly positive weights $A_{12}>0$ and $A_{23}>0$ such that $\bs \delta> 0$. \label{fig: existence}}
 \end{figure}

To avoid disconnecting edges or to maintain a fixed network topology,
a functional pattern should be realized in Problem \eqref{eq: problem
  2} using a strictly positive weight vector (that is,
$\bs \delta > 0$ rather than $\bs \delta \ge 0$). While, in
general, this is a considerably harder problem, a sufficient condition
for the existence of a strictly positive solution $\bs \delta>0$ is that 
the network with incidence matrix $\bar B$ contains an Hamiltonian path,
that is, a directed path that visits all the oscillators exactly once
(Fig.~\ref{fig: existence}\textbf{b} shows a network containing an
Hamiltonian path). Namely,

\emph{There exists a strictly positive solution $\bs \delta > 0$
  to
  $B D({\bs x}){\bs \delta} = \bs
    \omega$ if}
  \begin{itemize}
  \item[(ii.a)] \emph{the network  with incidence matrix $\bar B$
      contains a directed Hamiltonian path $\mc H$;}
  \item[(ii.b)] \emph{${ \bs \omega}^\transpose B_{:,i}D_{ii}  (\bs x)> 0$ for all $i\in \mc H$;}
  \end{itemize}

The incidence matrix $\bar B_{:,\mc H}$ of a directed Hamiltonian path $\mc H$
has two key properties. First, it comprises $n-1$ linearly independent
columns, since the path covers all the vertices and contains no
cycles. This guarantees that
$\bs \omega \in \mathrm{Im} (B_{:,\mc H})$. Second, the columns
of the incidence matrix $\bar B_{:,\mc H}$ satisfy
$\bar B_{:,i}^\transpose \bar B_{:,i} = 2$ and
$\bar B_{:,i}^\transpose \bar B_{:,j}\in\{0,-1\}$ for all
$i,j\in\mc H$, $i\ne j$. Then,
letting the set $\mc S$ in the result above identify the columns of
the Hamiltonian path, conditions (ii.a) and (ii.b) imply (i.a)-(i.c),
thus ensuring the existence of a nonnegative set of weights
$\bs \delta$ that solves the pattern assignment problem
$B D({\bs x}) \bs \delta =  { \bs
  \omega}$. Furthermore, 
  by rewriting the pattern assignment problem as in equation~\eqref{eq:solution-S},
the following vector of  interconnection weights solves such equation (see Methods):
\begin{align*}
  \bs  \delta_{\mc H} &= 
 \left(B_{:,\mc H} D_{\mc H,\mc H}({\bs x})\right)^\dagger \left( \bs \omega - B_{:,\tilde{\mc H}} D_{\tilde{\mc H},\tilde{\mc H}}({\bs x})\bs
  \delta_{\tilde{\mc H}} \right) .
\end{align*}
Because $\bar B_{:,\mc H} = B_{:,\mc H} D_{\mc H,\mc H}$ defines an
Hamiltonian path and because of (ii.b), the vector
$\left(B_{:,\mc H} D_{\mc H,\mc H}({\bs x})\right)^\dagger \bs \omega$
contains only strictly positive entries. Thus, for any sufficiently
small and positive vector $\bs\delta_{\tilde{\mc H}}$, the weights
$\bs\delta_{{\mc H}}$ are also strictly positive, ultimately proving the
existence of a strictly positive solution to the pattern assignment
problem (see Methods). Fig.~\ref{fig: existence}\textbf{c} illustrates
a self-contained example.}

Taken together, the results { presented in this section} reveal
that the interplay between the network structure and the oscillators'
natural frequencies dictates whether a desired functional pattern is
achievable under the constraint of { nonnegative (or even strictly
  positive) interconnections. First, dense positive networks with a large
  number of edges are more likely to generate a desired functional
  pattern, since their incidence matrix features a larger number of
  candidate vectors to satisfy conditions (i.a)-(i.c). Second, densely
  connected networks are also more likely to contain an Hamiltonian
  path, thus promoting also strictly positive network designs. Third,
  after an appropriate relabeling of the oscillators such that any
  interconnection from $i$ to $j$ in the Hamiltonian path satisfies
  $i<j$, condition (ii.b) is equivalent to requiring that
  $\omega_i<\omega_j$. That is, the feasibility of a functional
  pattern is guaranteed when the natural frequencies increase
  monotonically with the ordering identified by the Hamiltonian
  path. This also implies, for instance, that sparsely connected
  positive networks, and not only dense ones, can attain a large variety of
  functional patterns. An example is a connected line network with
  increasing natural frequencies, which can generate, among others,
  any functional pattern defined by phase differences that are smaller
  than $\frac{\pi}{2}$ (trivially, when the phase differences are
  smaller than $\frac{\pi}{2}$ and the natural frequencies are
  increasing, a line network contains an Hamiltonian path and the
  vector of natural frequencies has positive projections onto the
  columns of the incidence matrix). Fig. \ref{fig: mapping weights
    equilibria}{\bf a} contains an example of such a network.  }

\subsection*{Compatibility of multiple functional patterns}

{ A single choice of the interconnections weights can allow for
  multiple desired functional patterns, as long as they are compatible
  with the network dynamics in equation~\eqref{eq: kuramoto}. In this
  section, we provide a characterization of compatible
  functional patterns in a given network, and derive conditions for
  the existence of a set of interconnection weights that achieve multiple desired
  functional patterns.} Being able to concurrently assign multiple functional
  patterns is crucial, for instance, to the investigation and design
  of memory systems \cite{PSK-AA:20}, where different patterns of
  activity correspond to distinct memories. Furthermore, our results complement
  previous work on the search of equilibria in oscillator networks
  \cite{DM-NSD-FD-JDH:15}.

{ To find a set of functional patterns that exists concurrently in a
given network with fixed interconnection weights $\bs\delta$, we exploit the algebraic core of equation~\eqref{eq:
  linear} and show that the kernel of the incidence matrix $B$
uniquely determines the equilibria of the network. In fact, for a given network (i.e., $\bs\delta$ with nonzero components) all
compatible equilibria ${\bs x}^{(i)}$, $i\in\until{\ell}$, must
satisfy
\begin{equation}\label{eq: linear rewritten}
D(\bs x^{(i)}) \bs\delta = B^\dagger \bs\omega + \mathrm{ker}(B).
\end{equation}
From equation~\eqref{eq: linear rewritten}, we can see that the sine vector of all compatible equilibria must belong to a specific affine subspace of $\real^{|\mc E|}$:
\begin{equation}\label{eq: compatible x}
\sin(\bs x^{(i)}) \in \mathrm{diag}(\bs \delta)^{-1}\left( B^\dagger \bs\omega + \mathrm{ker}(B) \right).
\end{equation}
Rewriting equation~\eqref{eq: linear} in the above form connects the existence of distinct functional patterns with $\mathrm{ker}(B)$, the latter  featuring a number of well known properties. For instance, it holds that $\mathrm{dim}(\mathrm{ker}(B)) = |\mc E| - n + c$, where $c$ is the number of connected components in a network, and that $\mathrm{ker}(B)$ coincides with the subspace spanned by the signed path vectors of all undirected cycles in the network \cite{Godsil2001}. Notice also that, after a suitable reordering of the phase
differences in $\bs x$, we can write
$\sin({\bs x}) = [\sin(\bs
x_\mathrm{desired}^\transpose)~\sin(\bs
x_\mathrm{dep}^\transpose)]^\transpose$, where $\bs x_\mathrm{dep}$
denotes the phase differences dependent on $n-1$ desired phase
differences $\bs
x_\mathrm{desired}$. Thus, all the $\bs x_\mathrm{desired}$ for which 
 $\sin(\bs x_\mathrm{dep})$ intersects the
affine space described by $\mathrm{diag}(\bs \delta)^{-1}\left( B^\dagger \bs\omega + \mathrm{ker}(B) \right)$ identify compatible functional patterns.

To showcase how the intimate relationship between the network
structure and the kernel of its incidence matrix enables the
characterization of which (and how many) compatible patterns coexist,
we consider three essential network topologies: trees, cycles, and
complete graphs. For the sake of simplicity, we let $\bs\delta = \1$
and $\bs\omega = \bs 0$, so that equation~\eqref{eq: compatible x}
holds whenever $\sin(\bs x^{(i)}) \in \mathrm{ker}(B)$. In networks
with tree topologies it holds $\mathrm{ker}(B) = \bs 0$, and $\sin(\bs
x^{(i)})=\bs 0$ is satisfied by $2^{n-1}$ patterns of the form $x^{(i)}_{jk} = 0,\pi$, for all $(j,k)\in\mc E$. Consider now cycle networks, where $\mathrm{ker}(B) = \mathrm{span}~\1$. For any cycle of $n\ge 3$ oscillators, two families of patterns are straightforward to derive. First, there are $2^{n-1}$ patterns of the form $\bs x^{(i)}_{k,k+1} = 0,\pi$, with $k=1,\dots,n-1$, and $x^{(i)}_{n1} = -\sum_{k=1}^{n-1} x^{(i)}_{k,k+1}$. Second, there are $n-1$ splay states \cite{Doerfler2014}, where the oscillators' phases evenly span the unitary circle, with $ x^{(i)}_{jk} = \frac{2\pi m}{n}$, $m=1,\dots,n-1$, $(j,k)\in\mc E$. Fig.~\ref{fig: surfs 3
  nodes} illustrates the compatible functional patterns satisfying equation~\eqref{eq: compatible x} in a
positive network of three fully synchronizing oscillators.
\begin{figure}[t]
\centering
\includegraphics[width=.76\textwidth]{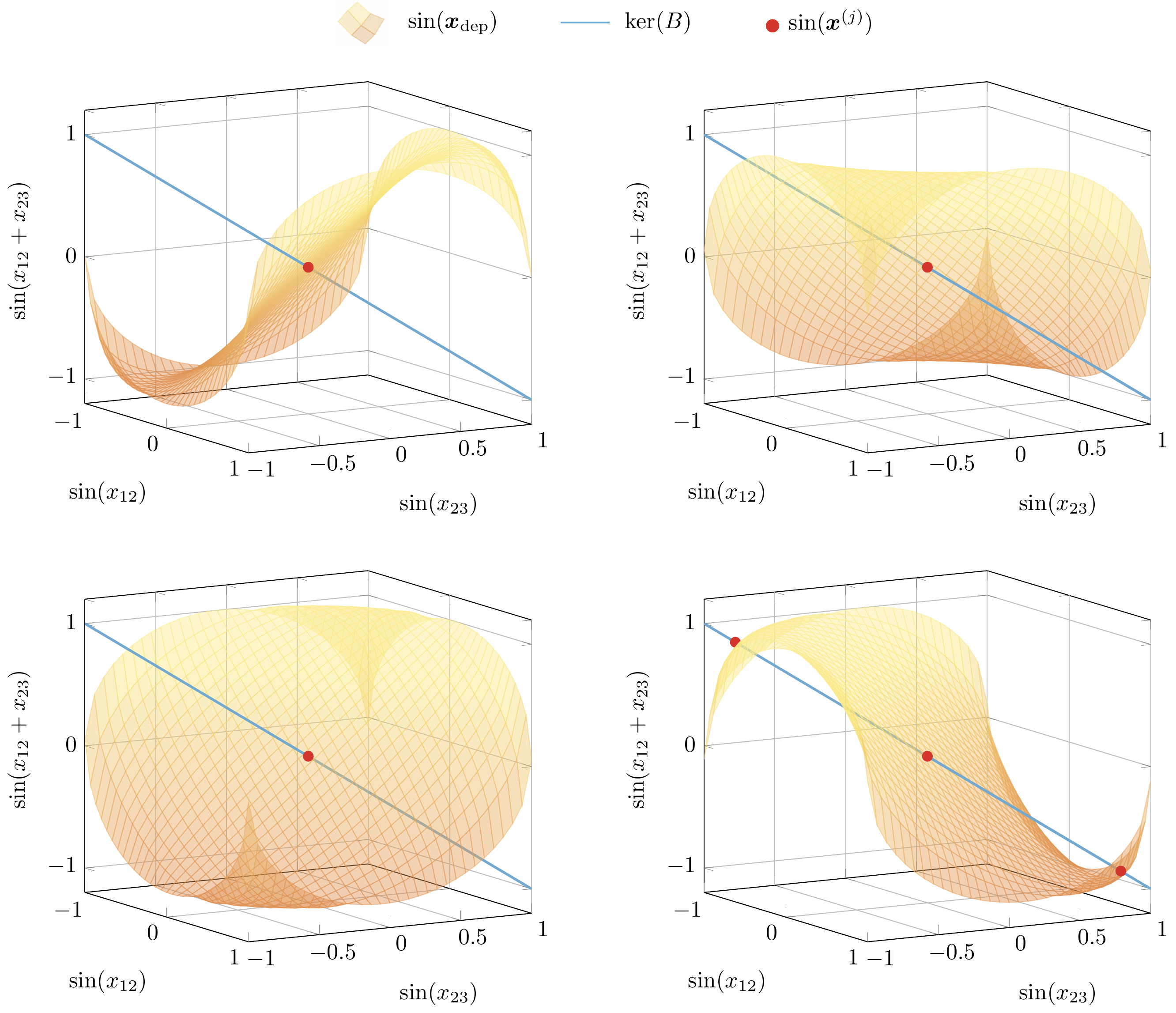}
\vspace{0.2cm}
\caption{ \textbf{The intersection of an affine space with $\sin(\bs x_\mathrm{dep})$ determines the compatible functional patterns of 3 identical oscillators.} Consider a fully connected network of $n=3$ identical oscillators with zero natural frequency and $\bs\delta=\1$. It is well known that ${\bs x}^{(1)} = [0~0]^\transpose$, ${\bs x}^{(2)} = [\pi~0]^\transpose$, ${\bs x}^{(3)} = [0~\pi]^\transpose$, ${\bs x}^{(4)} = [\pi~\pi]^\transpose$ are phase difference equilibria. Furthermore, because $\sin(\theta) = \sin(\pi-\theta)$, this figure illustrates $\sin(x_{13})$ as a function of $x_{12}$ and $x_{23}$ in four different panels: $\sin(x_{12}+x_{23})$ (top left), $\sin(\pi-x_{12}+x_{23})$ (top right), $\sin(x_{12}+\pi-x_{23})$ (bottom left), and $\sin(-x_{12}-x_{23})$ (bottom right). 
The fourth panel reveals that the two functional patterns compatible with $ {\bs x}^{(j)}$, $j\in\{1,\dots,4\}$, correspond to $ {\bs x}^{(5)} = [2\pi/3~2\pi/3]^\transpose$ and $ {\bs x}^{(6)} = [-2\pi/3~-2\pi/3]^\transpose$ (in red). \label{fig: surfs 3 nodes}}
\end{figure}
In general, however, cycle networks of identical oscillators admit infinite coexisting patterns. For instance, Fig.~\ref{fig: cycle patterns} shows how we can parameterize infinite equilibria with a scalar $\gamma\in\mathbb{S}^1$ in a cycle of $n=4$ oscillators. Finally, 
as complete graphs are equivalent to a composition of cycles, they also admit infinite compatible patterns that can be parameterized akin to what occurs in a simple cycle (see Supplementary Figure~\ref{fig: complete graph}).

We now turn our attention to finding the interconnection weights that simultaneously enable a collection of $\ell\ge 1$ desired functional patterns $\{{\bs x}^{(i)}\}_{i=1}^\ell$. We first notice that equation~\eqref{eq: linear rewritten} reveals that to achieve a desired functional pattern ${\bs x}^{(i)}$ with components not equal to $k\pi$, $k\in\mathbb{Z}$, the network weights $\bs\delta$ must belong to an $|\mc E|$-dimensional affine subspace of $\real^{|\mc E|}$:
\begin{equation}\label{eq: compatible weights}
 \bs\delta \in D(\bs x^{(i)})^{-1}\left( B^\dagger \bs\omega + \mathrm{ker}(B) \right), \quad \forall i=1,\dots,\ell.
\end{equation}
Let $\Gamma_i=D(\bs x^{(i)})^{-1}\left( B^\dagger \bs\omega +
  \mathrm{ker}(B) \right)$. Then, to concurrently realize a collection
of patterns $\{{\bs x}^{(i)}\}_{i=1}^\ell$, a solution to
equations~\eqref{eq: compatible weights} exists if and only if
$\bigcap_{i=1}^\ell \Gamma_i \ne \emptyset$. It is worth noting that,
whenever the latter intersection coincides with a singleton, then there exists a single choice of network weights that realizes 
$\{{\bs x}^{(i)}\}_{i=1}^\ell$. However, if $\bigcap_{i=1}^\ell \Gamma_i$ corresponds to a subspace, then infinite networks can realize the desired collection of functional patterns.
We conclude by emphasizing that a positive $\bs\delta$ that achieves
the desired patterns exists if and only if $\left(\bigcap_{i=1}^\ell
  \Gamma_i\right) \cap \real_{\ge 0}^{|\mc E|} \ne \emptyset$. That is, if the network weights belong to the nonempty intersection of the $\ell$ affine subspaces with the positive orthant.
    
    \begin{figure}[t]
\centering
\includegraphics[width=.9\textwidth]{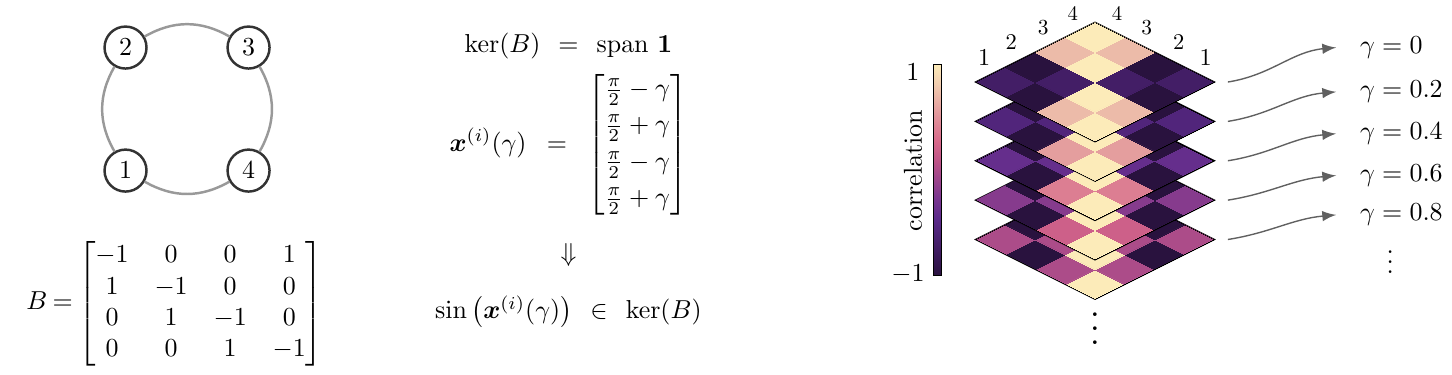}
\vspace{0.2cm}
\caption{ \textbf{A homogeneous cycle network admits infinite compatible functional patterns.} Since $\mathrm{ker}(B) = \mathrm{span}~\1$, the cycle network admits infinite compatible equilibria, which can be parameterized by $\gamma\in\mathbb{S}^1$ as $\bs x^{(i)}(\gamma) = [\frac{\pi}{2}-\gamma,~\frac{\pi}{2}+\gamma,~\frac{\pi}{2}-\gamma,~\frac{\pi}{2}+\gamma]^\transpose$. Any arbitrarily small variation of $\gamma$ yields $\sin(\bs x^{(i)}(\gamma)) \in \mathrm{ker}(B)$. The right panel illustrates the patterns associated with $\bs x^{(i)}(\gamma)$, $i=1,\dots,5$ for increments of $\gamma$ of 0.2 radians. \label{fig: cycle patterns}}
\end{figure}
}

\subsection*{Stability of functional patterns} 
{ A functional pattern is
stable when small deviations of the oscillators phases from the
desired configuration lead to vanishing functional perturbations.
Stability is a desired property since it guarantees that the desired
functional pattern is robust against perturbations to the oscillators
dynamics. To study the stability of a functional pattern, we analyze the
Jacobian of the Kuramoto dynamics at the desired functional
configuration, which reads as \cite{Doerfler2014}}
\begin{equation}\label{eq: Jacobian} J = \frac{\partial}{\partial \bs
    \theta}\dot{ \bs \theta} = -\underbrace{B
    \diag\left(\{A_{ij}\cos({
        x_{ij}})\}_{(i,j)\in\mc E}\right) B^\transpose}_{\mc
    L({ \bs x})},
\end{equation}
{ where $\mc L({ \bs x})$ denotes the Laplacian matrix of
the network with weights scaled by the cosines of the phase
differences (the weight between nodes $i$ and $j$ is
$A_{ij}\cos( \theta_j - \theta_i)$). The functional pattern
${\bs x}$ is stable when the eigenvalues of the above
Jacobian matrix have negative real parts. For instance, if all phase
differences are strictly smaller than $\frac{\pi}{2}$ (that is,
the infinity-norm of $\bs x$ satisfies $\| {\bs x} \|_\infty < \frac{\pi}{2}$), then the Jacobian
in equation~\eqref{eq: Jacobian} is known to be stable
\cite{Doerfler2014}.} In the case that both cooperative and competitive
interactions are allowed, we can ensure stability of a desired pattern
by specifying the network weights in $\bs \delta$ such that $A_{ij}>0$
if $|x_{ij}|<\frac{\pi}{2}$ and $A_{ij}<0$ otherwise, so that the
matrix $\mc L$ becomes the Laplacian of a positive network (see
Methods). { Furthermore}, we observe that in the particular case where some
differences $|x_{ij}|=\frac{\pi}{2}$, the network may become
disconnected since $\cos(x_{ij}) = 0$. { Because} the Laplacian of a
disconnected network has multiple eigenvalues at zero, marginal
stability may occur, and phase trajectories may not converge to the
{ desired} pattern.

{ When some phase differences are larger than $\frac{\pi}{2}$ and the
network allows only for nonnegative weights, then stability of a
functional pattern is more difficult to assess because the Jacobian
matrix becomes a \emph{signed} Laplacian \cite{DZ-MB:17}. The
off-diagonal entries of a signed Laplacian satisfy $\mc L_{ij}>0$
whenever $|x_{ij}|>\frac{\pi}{2}$, thus possibly changing the sign of
its diagonal entries $\mc L_{ii} = \sum_j A_{ij}\cos(x_{ij})$ and
violating the conditions for the use of classic results from algebraic
graph theory for the stability of Laplacian matrices. To derive a
condition for the instability of the Jacobian in equation \eqref{eq:
  Jacobian}, we exploit the notion of \emph{structural balance}.} We
say that the cosine-scaled network with Laplacian
matrix $\mc L$ is structurally balanced if and only if its oscillators
can be partitioned into two sets, $\mc O_1$ and $\mc O_2$, such that
all $(i,j)\in\mc E$ with $A_{ij}\cos(x_{ij})<0$ connect oscillators in
$\mc O_1$ to oscillators in $\mc O_2$, and all $(i,j)\in\mc E$ with
$A_{ij}\cos(x_{ij})>0$ connect oscillators within $\mc O_i$,
$i\in\{1,2\}$.  If a network is structurally balanced, then its
Laplacian has mixed eigenvalues \cite{DZ-MB:17}. Therefore, we
conclude the following:

\smallskip

\emph{ If the functional pattern $ {\bs x}$ yields a
  structurally balanced cosine-scaled network, then
  $ {\bs x}$ is unstable.}  \smallskip

The above condition allows us to { immediately} assess the
instability of functional patterns for the special cases of line and
cycle networks. { In fact, for a line network with positive
  weights, $\bs x$ is unstable whenever $|x_{ij}|>\frac{\pi}{2}$ for
  any $i,j$. Instead, for a cycle network with positive weights, the
  pattern $\bs x$ can be stable only if it contains at most one phase
  difference $\frac{\pi}{2}<|x_{ij}|< \gamma$, where $\gamma \approx 1.789776$ solves $\gamma-\tan(\gamma)=2\pi$ (see Supplementary
  Information). In the next section, we propose a heuristic procedure
  to correct the interconnection weights in positive networks to
  promote stability of a functional pattern.}

\subsection*{Optimal interventions for desired functional patterns}
Armed with conditions to guarantee the existence of positive
interconnections that enable a desired functional pattern, we { now
  show that the problem of adjusting the} network weights to generate
a desired functional pattern can be cast as a convex optimization
problem. Formally, for a { desired functional pattern} ${\bs x}$
and network weights $\bs \delta$, we seek to solve
\begin{align}\label{eq: optimization}
\min_{\bs \alpha} &&&  \left\| \bs \alpha \right\|_{2} \\
\mathrm{subject~to }&&& B D({\bs x}) (\bs \delta + \bs \alpha) ={\bs \omega}, \tag{\theequation a}\label{eq: constraint} \\
\text{ and }&&& (\bs \delta + \bs \alpha) \ge 0, \tag{\theequation b}\label{eq: positive constraint}
\end{align}
{ where $\bs \alpha\in\real^{|\mc E|}$ are the controllable
  modifications of the network weights, and $\|\cdot\|_{2}$ denotes
  the $\ell^2$-norm. Fig.~\ref{fig: control examples}{\bf a} illustrates the control of a functional pattern in a line network of $n=4$ oscillators.

\begin{figure}[t]
  \centering
  \includegraphics[width=.95\textwidth]{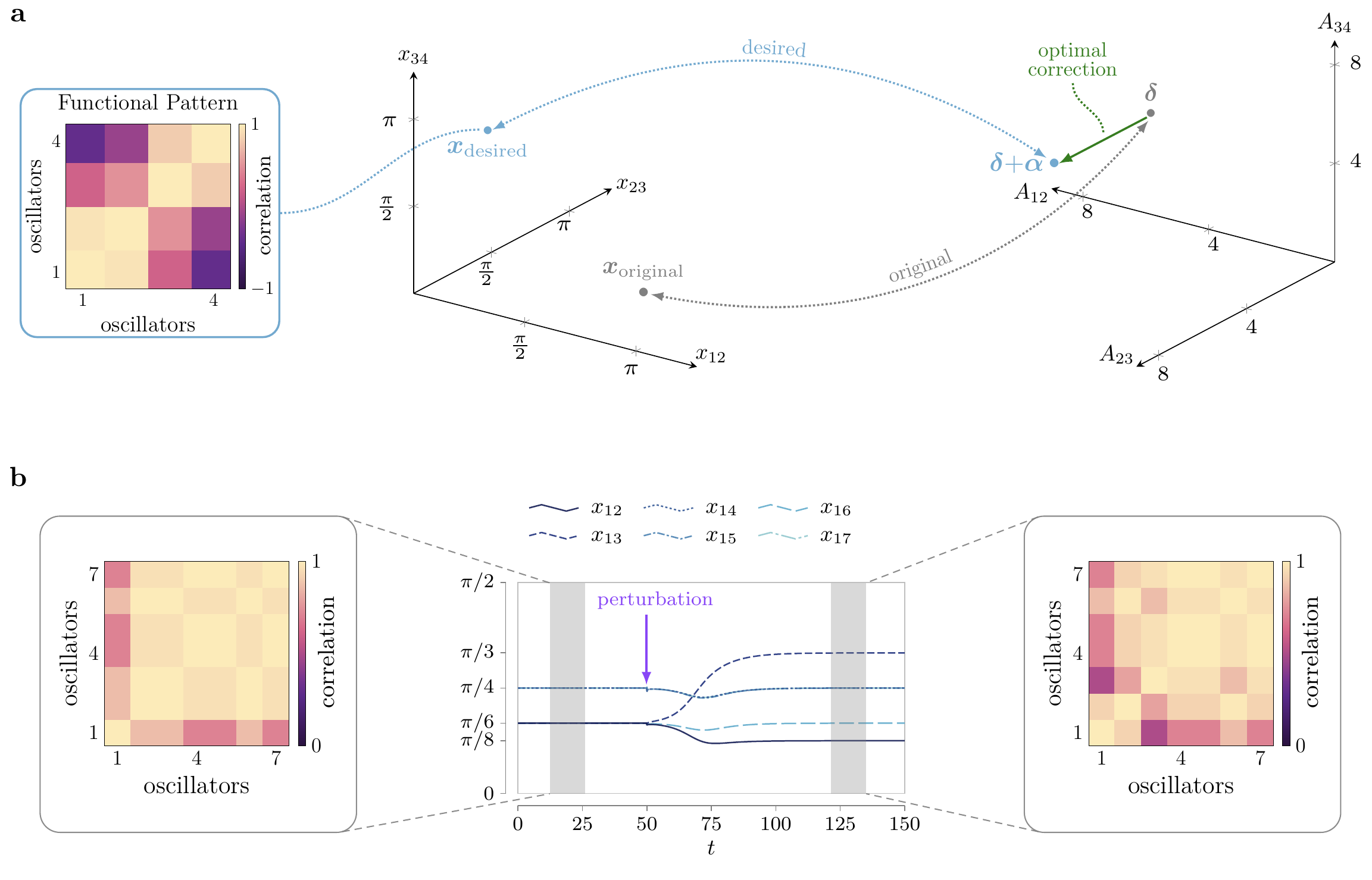}
  \vspace{0.2cm}
  \caption{\textbf{Optimal interventions for desired functional
      patterns.} {\bf a} For the line network in Fig.~\ref{fig: mapping weights
      equilibria}{\bf a}, we solve Problem~\eqref{eq: optimization} to
    assign the desired pattern
    $\bs x_\text{desired} =
    [\frac{4\pi}{5}~\frac{\pi}{3}~\frac{\pi}{10}]^\transpose$. The
    starting pattern
    $\bs x_\text{original} =
    [\frac{\pi}{10}~\frac{\pi}{3}~\frac{4\pi}{5}]^\transpose$ is
    associated to interconnection weights
    $\bs \delta = [3.4026~3.4641~6.4721]^\transpose$. Applying the
    optimal correction $\bs \alpha^*$ yields positive interconnection
    weights
    $\bs\delta + \bs \alpha^* = [6.4721~3.4026~3.4641]^\transpose$ that
    achieve the desired functional patterns $\bs
    x_\text{desired}$. {\bf b} Joint allocation of two compatible equilibria for the phase difference dynamics. By taking $\theta_1$ as a reference, we choose two points for the phase differences $x_{1i} = \theta_i-\theta_1$, $i\in\fromto{2}{7}$, to be imposed as equilibria in a network of $n=7$ oscillators: ${\bs x}^{(1)}_\text{desired} = \left[\frac{\pi}{6}~ \frac{\pi}{6}~ \frac{\pi}{4}~ \frac{\pi}{4}~ \frac{\pi}{6}~ \frac{\pi}{4}\right]^\transpose$ and ${\bs x}^{(2)}_\text{desired} = \left[\frac{\pi}{8}~ \frac{\pi}{3}~ \frac{\pi}{4}~ \frac{\pi}{4}~ \frac{\pi}{6}~\frac{\pi}{4}\right]^\transpose$. In this example, we find a set of interconnection weights $(\bs\delta + \bs\alpha^*)$ that solves the minimization problem~\eqref{eq: optimization} with constraint~\eqref{eq: multiple control}.
The trajectories start at the (unstable) equilibrium point ${\bs x}^{(1)}_\text{desired}$ at time $t=0$, and converge to the point ${\bs x}^{(2)}_\text{desired}$ after a small perturbation $\bs p\in\mathbb{T}^7$, with $p_i\in[0~0.05]$, is applied to the phase difference trajectories at time $t=50$.
    \label{fig: control examples}}
\end{figure}

  The minimization problem \eqref{eq: optimization} is convex, thus
  efficiently solvable even for large networks, and may admit multiple
  minimizers, thus showing that different networks may exhibit the
  same functional pattern. Moreover, in light of our results above, Problem \eqref{eq: optimization} can be easily adapted to assign a collection of desired patterns $\{\bs x^{(i)}\}_{i=1}^\ell$. To do so, we simply replace the constraint~\eqref{eq: constraint} with
  \begin{equation}\label{eq: multiple control}
  BD(\bs x^{(i)}) (\bs\delta + \bs \alpha) = \bs \omega,\quad \forall i=1,\dots,\ell.
  \end{equation} 
  Fig.~\ref{fig: control examples}{\bf b} illustrates an example where we jointly
    allocate two functional patterns for a complete graph with $n=7$
  oscillators (see Supplementary Information for more details on this
  example).
  
  Note that the minimization problem
  \eqref{eq: optimization} does not guarantee that the functional
  pattern $\bs x$ is stable for the network with weights $\bs \delta +
  \bs \alpha^*$. To promote stability of the pattern $\bs x$, we use a
  heuristic procedure based on the classic Gerschgorin's theorem
  \cite{Gerschgorin1931}. Recall that stability of $\bs x$ is
  guaranteed when the associated Jacobian matrix has a Laplacian
  structure, with negative diagonal entries and nonnegative
  off-diagonal entries. Further, instability of $\bs x$ depends
  primarily on the negative off-diagonal entries $A_{ij}\cos(
  x_{ij})$  of the Jacobian (these entries are negative because the
  sign of the network weight $A_{ij}$ is different from the sign of
  the cosine of the desired phase difference $x_{ij}$). Thus, reducing
  the magnitude of such entries $A_{ij}$ heuristically moves the
  eigenvalues of the Jacobian towards the stable half of the complex
  plane (this phenomenon can be captured using the Gerschgorin circles, as we show in
  Fig.~\ref{fig: eig shift} for a network with $7$ nodes). To
  formalize this procedure, let $\bs \delta_{\mc N}$ and $\bs
  \alpha_{\mc N}$ denote the entries of the weights $\bs\delta$ and
  tuning vector $\bs \alpha$, respectively, that are associated to negative interconnections $A_{ij}\cos(
  x_{ij})<0$ in the cosine-scaled network. Then, the optimization problem that enacts the proposed strategy becomes:
\begin{align}\label{eq: heuristic}
\min_{\bs  \alpha} &\ \ \left\| \bs\delta_{\mathcal N} + \bs \alpha_{\mathcal N} \right\|_{2}\\
\mathrm{subject~to }&\ \ B D({\bs x}) (\bs \delta + \bs\alpha) = {\bs  \omega}, \notag\\
\mathrm{and}&\ \ (\bs \delta + \bs \alpha) \ge 0 \notag,
\end{align}

Carefully reducing the weights $\bs\delta_{\mathcal N}+\bs\alpha_{\mathcal N}$
promotes stability of the target pattern. Fig.~\ref{fig: eig
  shift} illustrates the shift of the Jacobian's eigenvalues while the
optimal tuning vector $\bs\alpha^*$ is gradually applied to a
$7$-oscillator network to achieve stability of a functional pattern
containing negative correlations (the network parameters can be found in
the Supplementary Information). 
Finally, we remark that the procedure in \eqref{eq: heuristic} can be further refined by introducing scaling constants to penalize $\left\| \bs\delta_{\mathcal N} + \bs \alpha_{\mathcal N} \right\|_{2}$ differently from the modification of other interconnection weights (see Supplementary Information for further details and an example).

\begin{figure}[t]
\centering
\includegraphics[width=1\textwidth]{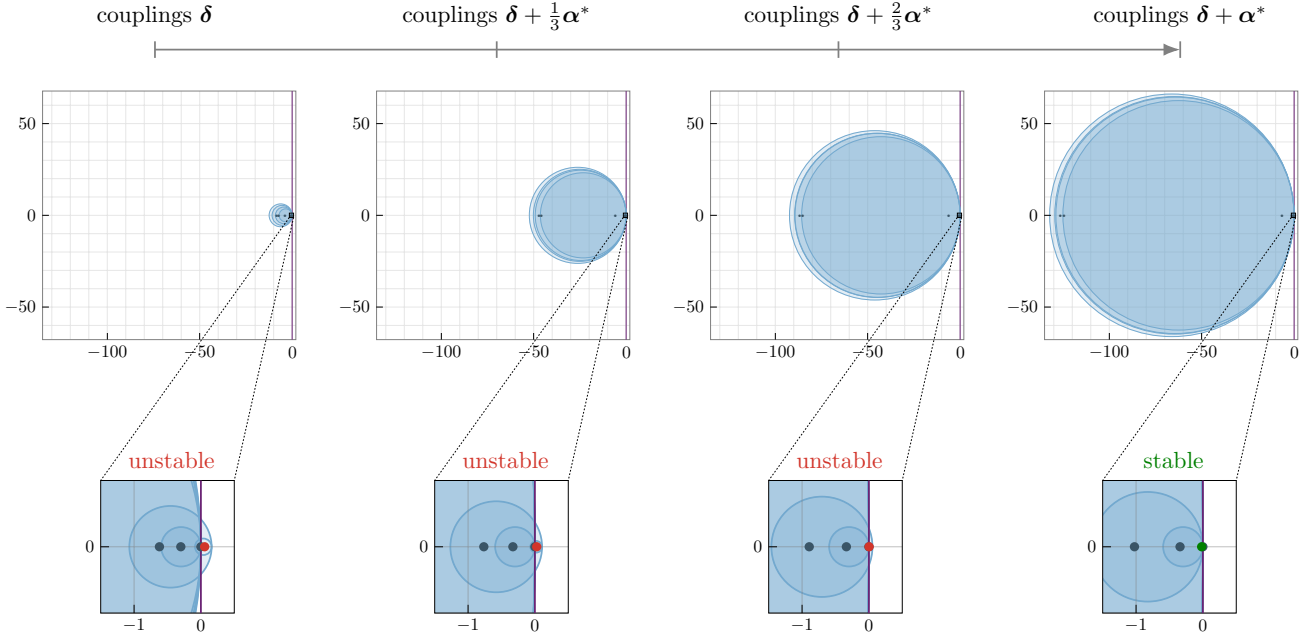}
 \vspace{0.4cm}
 \caption{ \textbf{Mechanism underlying the heuristic procedure to
     promote stability of functional patterns containing negative
     correlations.} For the $7$-oscillator network in Supplementary
   Text 1.5, we apply the procedure in equation~\eqref{eq: heuristic}
    to achieve stability of the pattern
   ${\bs x}_\text{desired} = \left[\frac{21\pi}{32} ~ \frac{\pi}{6}
     ~\frac{\pi}{6} ~\frac{\pi}{8} ~\frac{\pi}{8}
     ~\frac{\pi}{3}\right]^\transpose$, where
   $x_{12} = \theta_2-\theta_1 >\frac{\pi}{2}$. The left plot
   illustrates the Gerschgorin disks (in blue) and the Jacobian's eigenvalues
   locations for the original network (as dark dots). The complex axis is highlighted in purple. It can be observed in the
   zoomed-in panel that one eigenvalue is unstable
   ($\lambda_2 = 0.0565$, in red). The optimal correction
   $\bs\alpha^*$ is gradually
   applied to the existing interconnections from the left-most panel to the right-most one at
   $\frac{1}{3}$ increments. The right zoomed-in panel shows that, as
   a result of our procedure, $n-1$ eigenvalues ultimately lie in the
   left-hand side of the complex plane ($\lambda_1=0$ due to
   rotational symmetry and $\lambda_2=-0.0178$, in green). \label{fig:
     eig shift}}
\end{figure}
}

{ The minimization problems \eqref{eq: optimization} and \eqref{eq: heuristic} do not
allow us to tune the oscillators' natural frequencies, and are constrained to networks with positive weights. When any parameter of the network is unconstrained and can be adjusted to enforce a desired functional
pattern, the network optimization problem can be generalized as}
\begin{align}\label{eq: optimization 2}
\min_{\bs \alpha, \bs \beta} &&&  \left\| \begin{bmatrix}\bs \alpha^\transpose & \bs \beta^\transpose \end{bmatrix}\right\|_{2} \\
\mathrm{subject~to }&&& B D({\bs x}) (\bs \delta + \bs \alpha) ={[\omega_1~\cdots~\omega_n]^\transpose} + \bs \beta, \tag{\theequation a}\label{eq: constraint 2}
\end{align}
{ where $\bs\beta$ denotes the correction to the natural frequencies. Problem~\eqref{eq: optimization 2} always admits a
solution because $\bs\beta$ can be chosen to satisfy the constraint
\eqref{eq: constraint 2} for any choice of the network parameters
$\bs\delta+\bs\alpha$. Further, the (unique)} solution to the minimization
problem \eqref{eq: optimization 2} can also be computed  in closed form:
\begin{align*}
\begin{bmatrix} \bs\alpha^* \\ \bs\beta^*\end{bmatrix} = \begin{bmatrix} B D( {\bs x}) & { -}I_n \end{bmatrix}^\dagger \left({[\omega_1~\cdots~\omega_n]^\transpose} - B D( {\bs x}) \bs \delta\right)
\end{align*}
where $I_n$ denotes the $n\times n$ identity matrix.

{ We conclude this section by noting that the minimization problems
  \eqref{eq: optimization}-\eqref{eq: optimization 2} can be
  readily extended to include other vector norms besides the
  $\ell_2$-norm in the cost function (e.g., the $\ell_1$-norm to
  promote sparsity of the corrections), and to be applicable to directed networks. The latter extension can be attained by replacing the
  constraints \eqref{eq: constraint} and \eqref{eq: constraint 2} with
  a suitable rewriting of the matrix form \eqref{eq: linear}. We refer
  the interested reader to the Supplementary Information for a
  comprehensive treatment of this extension and an example. }

\section*{Applications to complex networks}

{ In the remainder of this paper we apply our methods to an
  empirically reconstructed brain network and to the IEEE 39 New
  England power distribution network. In the former case, we use the
  Kuramoto model to map structure to function, and find that local
  metabolic changes underlie the emergence of functional patterns of
  recorded neural activity. In the latter case, we use our methods to
  restore the nominal network power flow after a fault.}

\subsection*{Local metabolic changes govern the emergence of distinct functional patterns in the brain}

The brain can be studied as a network system in which Kuramoto
oscillators approximate the rhythmic activity of different brain
regions
\cite{Corbetta2015,AP-MR:15,JC-EH-OS-GD:11,TM-GB-DSB-FP:19b}. Over
short time frames, the brain is capable of exhibiting a rich
repertoire of functional patterns while the network structure and the
interconnection weights remain unaltered. Functional patterns of brain
activity not only underlie multiple cognitive processes, but can also
be used as biomarkers for different psychiatric and neurological
disorders \cite{AS-JG:05}.

To shed light on the structure-function relationship of the human
brain, we utilize Kuramoto oscillators evolving on an empirically
reconstructed brain network. We hypothesize that the intermittent
emergence of diverse patterns stems from changes in the oscillators'
natural frequencies -- which can be thought of as endogenous changes
in metabolic regional activity regulated by glial cells
\cite{RDF-et-al:14} or exogenous interventions to modify undesired
synchronization patterns \cite{TM-GB-DSB-FP:19b}. First, we show that
phase-locked trajectories of the Kuramoto model in equation~\eqref{eq:
  kuramoto} can be accurately extracted from noisy measurements of
neural activity and are a relatively accurate approximation of neural
activity.

We employ structural (i.e., interconnections between brain regions)
and functional (i.e., time series of recorded neural activity) data
from Ref.~\cite{Corbetta2015}. Structural connectivity consists of a
sparse weighted matrix $A$ whose entries represent the strength of the
physical interconnection between two brain regions. Functional data
comprise time series of neural activity recorded through functional
magnetic resonance imaging (fMRI) of healthy subjects as they
rest. Because the phases of the measured activity have been shown to
carry most of the information contained in the slow oscillations
recorded through fMRI time series, we follow the steps in
Ref.~\cite{Corbetta2015} to obtain such phases from the data by
filtering the time series in the $[0.04, 0.07]$ Hz frequency range
(Supplementary Information). Next, since frequency synchronization is
thought to be a crucial enabler of information exchange between
different brain regions and homeostasis of brain dynamics
\cite{FV-JPL-ER-JM:01,RN-PD-JDB-SB-PWFDM-JL:20}, we focus on
functional patterns that arise from phase-locked trajectories, as
compatible with our analysis. For simplicity, we restrict our study to
the cingulo-opercular cognitive system, which, { at the spatial
  scale of our data,} comprises $n=12$ interacting brain regions
\cite{JDP-others:11}.

\begin{figure}[t]
\centering
 \includegraphics[width=.9\columnwidth]{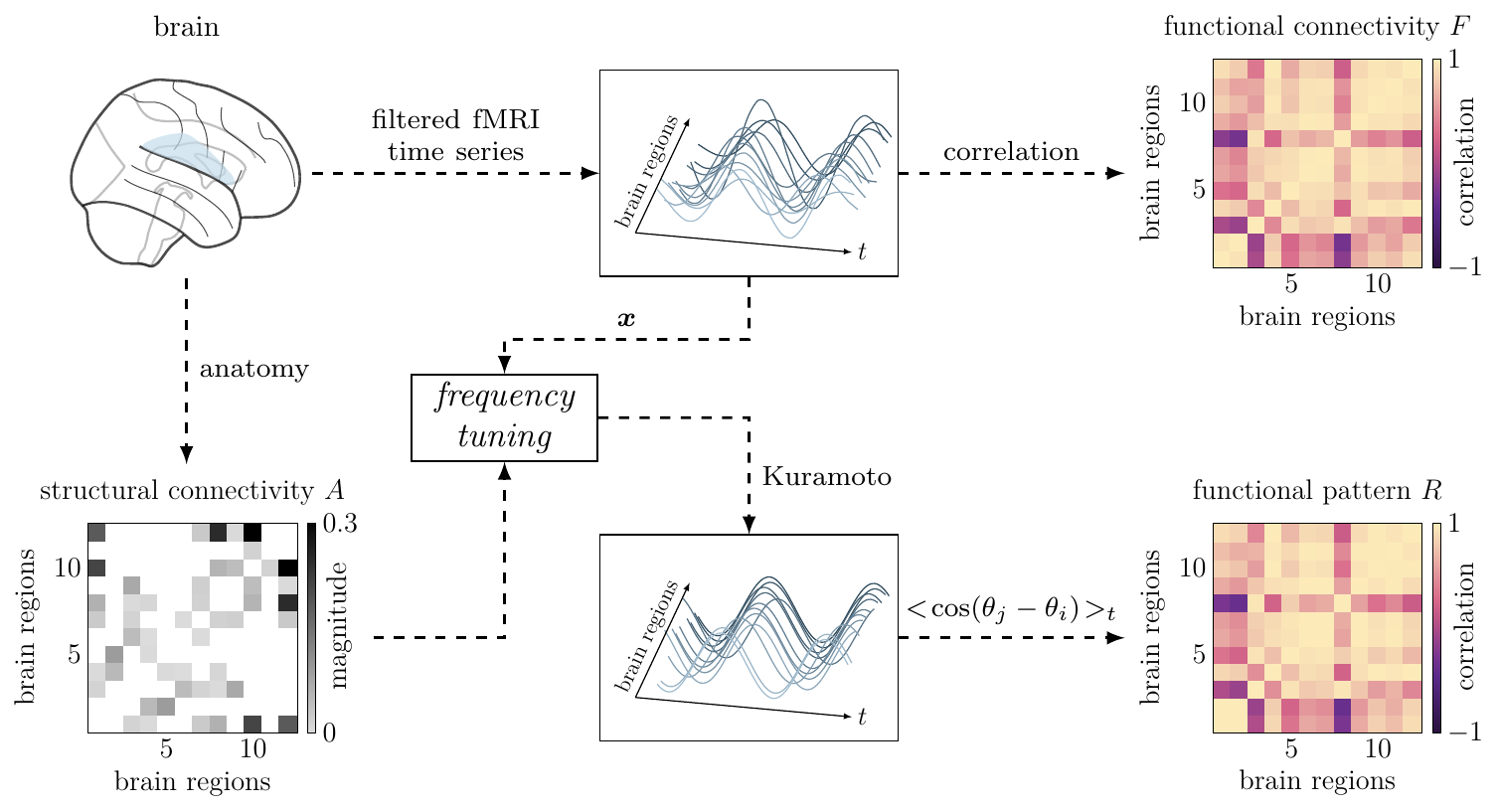}%
 \vspace{0.2cm}
 \caption{\textbf{Replication of empirically recorded functional
     connectivity in the brain through tuning of the natural
     frequencies of Kuramoto oscillators.} The anatomical brain
   organization provides the network backbone over which the
   oscillators evolve. The filtered fMRI time series provide the
   relative phase differences between co-fluctuating brain regions,
   and thus define the desired phase differences $ {\bs x}$, which is used to
   calculate the metabolic change encoded in the oscillators' natural
   frequencies. In this figure, we select the $40$-second time window
   from $t_0 = 498$ seconds to $t_f = 538$ seconds for subject $18$ in
   the second scanning session. We obtain $\| R - F\|_2 =
   0.2879$. Additionally, we verify that the analysis of the Jacobian
   spectrum (see equation~\eqref{eq: Jacobian}) accurately predicts
   the stability of the phase-locked trajectories. { Supplementary
     Figure 7{\bf a} illustrates the basin of attraction of $R$, which
     we numerically estimate to be half of the torus.} \label{fig:
     brain}}
 \end{figure}
 
 We identify time windows in the filtered fMRI time series where the
 signals are phase-locked, and compute two matrices for each time
 window: a matrix $F\in\real^{12\times 12}$ of Pearson correlation
 coefficients (also known as functional connectivity), and a
 functional pattern $R\in\real^{12\times 12}$ (as in
 equation~\eqref{eq: corr coeff}) from the phases extracted by solving
 the \emph{nonconvex phase synchronization} problem
 \cite{NB:16}. Strikingly, we find that $\|F-R\|_2\ll 1$ consistently
 (see Supplementary Information and Supplementary Figure~\ref{fig: FC-R}\textbf{b}),
 thus demonstrating that our definition of functional pattern is a
 good replacement of the classical Pearson correlation arrangements in
 networks with oscillating states.

 Building upon the above finding, we test whether the oscillators'
 natural frequencies embody the parameter that links the brain network
 structure to its function (i.e., structural and functional
 matrices). We set
 $ \bs \omega = B D({ \bs x}) \bs \delta$, where
 ${ \bs x}$ are phase differences obtained from the
 previous step, and integrate the Kuramoto model in
 equation~\eqref{eq: kuramoto} with random initial conditions close to
 ${ \bs x}$. We show in Fig.~\ref{fig: brain} that
 the assignment of natural frequencies according to the extracted
 phase differences promotes spontaneous synchronization to accurately
 replicate the empirical functional connectivity $F$.

 These results corroborate the postulate that structural connections
 in the brain support the intermittent activation of specific
 functional patterns during rest through regional metabolic
 changes. Furthermore, we show that the Kuramoto model represents an
 accurate and interpretable mapping between the brain anatomical
 organization and the functional patterns of frequency-synchronized
 neural co-fluctuations.  Once a good mapping is inferred, it can be
 used to define a generative brain model to replicate \emph{in silico}
 how the brain efficiently enacts large-scale integration of
 information, and to develop personalized intervention schemes for
 neurological disorders related to abnormal synchronization phenomena
 \cite{FV-MS-PJH-GS-JC-RL:15,TM-GL-FP-AC:20}.

\subsection*{Power flow control in power networks for fault recovery and prevention}
Efficient and robust power delivery in electrical grids is crucial for
the correct functioning of this critical infrastructure. Modern,
reconfigurable power networks are expected to be resilient to
distributed faults and malicious cyber-physical attacks
\cite{JWW-LLR:09} while being able to rapidly adapt to varying load
demands. { In addition, climate change is straining service
  reliability, as underscored by recent events such as the Texas grid
  collapse after Winter Storm Uri in February 2021
  \cite{JWB-KB-MDB-AQG-EG-VR-JDR-SS-CAS-MEW:21}, and the New Orleans
  blackout after Hurricane Ida in August 2021 \cite{RS-JR:21}.}
Therefore, there exists a dire need to design control methods to
efficiently operate these networks and react to unforeseen disruptive
events.
 
The Kuramoto model in equation~\eqref{eq: kuramoto} has been shown to
be particularly relevant in the modeling of large distribution
networks and microgrids \cite{FD-MC-FB:13}. Preliminary work on the
control of frequency synchronization in electrical grids modeled
through Kuramoto oscillators has been developed in
Ref.~\cite{PSS-AA:15}. Here, we present a method that leverages our
findings to guarantee not only frequency synchronization, but also a
{ target} pattern of active power flow. Our method can be used for
power (re)distribution with respect to specific pricing strategies,
fault prevention (e.g., when a line overheats) and recovery (e.g.,
after the disconnection of a branch).  { Furthermore, thanks to the
  formal guarantees that our method prescribes, we are able to prevent
  Braess' Paradox in power networks \cite{DW-MT:12}, which is a
  phenomenon where the addition of { interconnections} to a network
  may impede its synchronization.}

It has been shown in Ref.~\cite[Lemma 1]{FD-MC-FB:13} that the load
dynamics (nodes $1$-$29$ in Fig.~\ref{fig: power network}\textbf{a})
in a structure-preserving power grid model have the same stable
synchronization manifold of equation~\eqref{eq: kuramoto}. In this
model, { $\omega_i = p_{\ell_i}/D_i$ is the active power load at
  node $i$, where $D_i$ is the damping coefficient}, and
$A_{ij} = |v_i||v_j|\mathcal{I}({Y}_{ij})/D_i$, with $v_i$ denoting
the nodal voltage magnitude and $\mathcal{I}({Y}_{ij})$ being the
imaginary part of the admittance $Y_{ij}$ {(see Supplementary
  Information for details about this model). In this example, we choose
  a highly damped scenario where $D_i=1$, which is possibly due to
  local excitation controllers.}  Notice that, when the phase angles
$ \bs \theta$ are phase-locked and $A_{ij}$ is fixed, the active power
flow is given by $A_{ij}\sin(\theta_j-\theta_i)$.
 
We posit that solving the problem in equation~\eqref{eq: optimization}
to design a local reconfiguration of the network parameters can
recover the power distribution before a line fault or provide the
smallest parameter changes to steer the load powers to desired
values. In practice, control devices such as Flexible Alternating
Current Transmission Systems (FACTs) allow operators and engineers to
change the network parameters
\cite{AL-GM-UH-CR-SR-HF-GF-AP:10,MAE-MSA-AM-NT:13}. We demonstrate the
effectiveness of our approach by recovering a desired power
distribution in the IEEE 39 New England power distribution network
after a fault. During a regime of normal operation, we simulate a
fault by disconnecting the line between two loads and solve the problem
in equation~\eqref{eq: optimization} to find the minimum modification
of the couplings $a_{ij}$ that recovers the nominal power
distribution.

  \begin{figure*}
\centering
 \includegraphics[width=1\columnwidth]{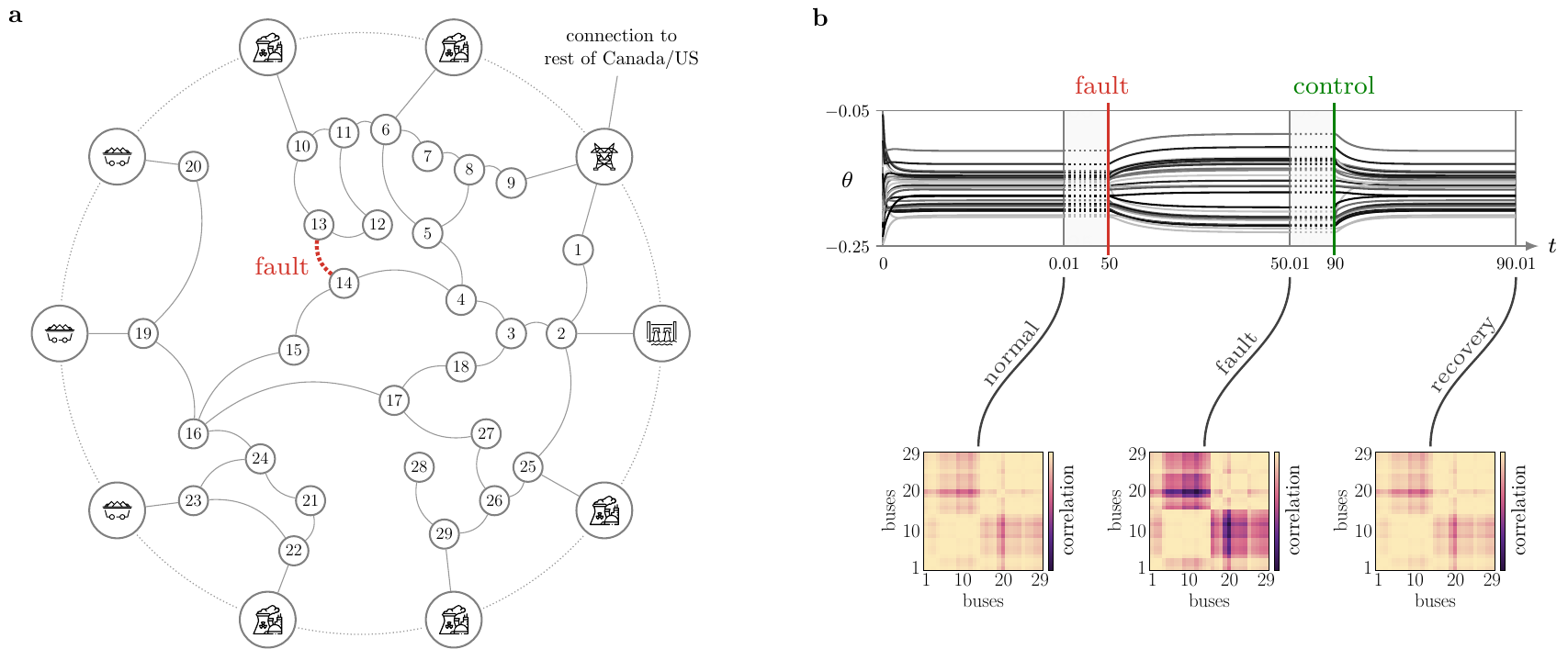}%
  \vspace{0.3cm}
  \caption{\textbf{Fault recovery in the IEEE 39 New England power
      distribution network through minimal and local intervention.}
    \textbf{a} New England power distribution network. The generator
    terminal buses illustrate the type of generator (coal, nuclear,
    hydroelectric). We simulate a fault by disconnecting the
    transmission line $25$ (between loads $13$ and $14$). \textbf{b}
    The fault causes the voltage phases $\bs \theta$ to depart from
    normal operating conditions, which could cause overheating of some
    transmission lines (due to violation of the nominal thermal
    constraint limits) or abnormal power delivery. To recover the
    pre-fault active power flow and promote a local (sparse)
    intervention, we solve the optimization in equation~\eqref{eq:
      optimization} by minimizing the $1$-norm of the structural
    parameter modification $ \bs \delta$ { with no scaling
      parameters in the cost functional}. The network returns to the
    initial operative conditions with a localized modification of the
    neighboring transmission lines' impedances. \label{fig: power
      network}}
 \end{figure*}
 
 We first utilize MATPOWER \cite{RDZ-CEMS-RJT:10} to solve the power
 flow problem. Then we use the active powers $ \bs p_{\ell}$ and
 voltages $ v$ at the buses to obtain the natural frequencies
 ${ \bs \omega}$ and the adjacency matrix $A$ of the oscillators,
 respectively, while the voltage phase angles are used as initial
 conditions $ \bs \theta(0)$ for the Kuramoto model in
 equation~\eqref{eq: kuramoto}. We integrate the Kuramoto dynamics and
 let the voltage phases $\bs \theta(t)$ reach a frequency-synchronized
 steady state, which corresponds to a normal operating condition. The
 phase differences also represent a functional pattern across the
 loads. Next, to simulate a line fault, we disconnect one line. By
 solving the problem in equation~\eqref{eq: optimization} with
 ${ \bs x}_\mathrm{desired}$ corresponding to the pre-fault
 steady-state voltage phase differences, we compute the smallest
 variation of the remaining parameters (i.e., admittances) so that the
 original functional pattern can be recovered.  Fig.~\ref{fig: power
   network}\textbf{b} illustrates the effectiveness of our procedure
 at recovering the nominal pattern of active power flow by means of
 minimal and localized interventions (see also Supplementary Figure~\ref{fig: modification}).

 { The above application is based on a classical lossless
   structure-preserving power network model
   \cite{FD-MC-FB:13}. However, in the power systems literature, more
   complex dynamics that relax some of the modeling assumptions have
   been proposed. For instance, Ref.~\cite{KS-LRG-MM-TF-DW:18} uses a
   third-order model (or one-axis model) that takes into account
   transient voltage dynamics. Instead,
   Ref.~\cite{FH-PS-RL-TK-JK-YM:20} studies the case of
   interconnections with power losses, which lead to a network of
   phase-lagged Kuramoto-Sakaguchi oscillators \cite{HS-YK:86}. We
   show in the Supplementary Information that our procedure to recover
   a { target} functional pattern can still be applied successfully
   to a wide range of situations involving these more realistic
   models.}

\section*{Contribution and future directions}

Distinct configurations of synchrony govern the { functioning} of
oscillatory network systems. This work presents a simple and
mathematically grounded mapping between the structural parameters of
arbitrary oscillator networks and their components' functional
interactions. { The tantalizing idea of prescribing patterns in
  networks of oscillators has been investigated before, yet only
  partial results had been reported in the literature. Here, we}
demonstrate that the control of patterns of synchrony can be cast as
optimal (convex) design and tuning problems. We also investigate the
feasibility of such optimizations in the cases of networks that admit
negative coupling weights and networks that are constrained to
positive couplings. Our control framework also allows us to prescribe
multiple desired equilibria in Kuramoto networks, a problem that is
relevant in practice and had not been investigated before.

{ As stability of a functional pattern may be a compelling property
  in many applications, we explore conditions to test and enforce the
  stability of functional patterns. We show that such conditions are
  rather straightforward in the case of networks that admit both
  cooperative and competitive interactions. However, stability of
  target functional patterns in networks that are constrained to only
  cooperative interactions is a more challenging task, for which we
  demonstrate that any pattern associated to a structurally-balanced
  cosine-weighted network cannot be stable. To overcome this issue, we
  propose an heuristic procedure that adjusts the oscillators'
  coupling strengths to violate the structurally-balanced property and
  promote the stability of functional patterns with negative
  correlations. While heuristic, our procedure for stability has
  proven successful in all our numerical studies. Notice that, differently from methods that study an ``average" description of the system at near-synchronous state (see, e.g., Ref.~\cite{JS-EMB-TN:09}), here we assess the stability of exact target phase-locked trajectories where phase values can instead be arbitrarily spread over the torus. This method can also be extended to assess the stability of equilibria of higher-dimensional oscillators, provided that the considered equilibrium state is a fixed point.}

We { emphasize} that our results are also intimately related to the
long standing economic problem of enhancing network operations while
optimizing wiring costs. In any complex system where synchrony between
components ensures appropriate functions, it is beneficial to maximize
synchronization while minimizing the physical variations of the
interconnection weights \cite{TN-AEM:06}. Compatible with this
principle, neural systems are thought to have evolved to maximize
information processing by promoting synchronization through optimal
spatial organization \cite{SA-EB:07}. Inspired by the efficiency
observed in neurobiological circuitry, equation~\eqref{eq:
  optimization} could be utilized for the design optimal interaction
schemes in large-scale computer networks whose performance relies on
synchronization-based tasks \cite{GK-MAN-HG-ZT-PAR:03}.

{ An important consideration that highlights the general
  contributions of the present study is that being able to specify
  pairwise functional relationships between the oscillators also
  solves problems such as phase-locking, full, and cluster
  synchronization. Yet, the converse is not true. In fact, even in the
  general setting of cluster synchronization \cite{Pecora2014} --
  where distinct groups of oscillators behave cohesively -- one can
  only achieve a desired synchronization level \emph{within} the same
  cluster, but not \emph{across} clusters, which instead is possible
  with our approach. Specifically, in cluster-synchronization
  oscillators belonging to the same cluster} are forced to
synchronize, thus implying that the associated diagonal blocks of the
functional pattern $R$ display values close to $1$. Seminal work in
Ref.~\cite{IK-CGR-HK-JLH:07} developed a nonlinear feedback control to
change the coupling functions in equation~(2) to engineer clusters of
synchronized oscillators, whereas the authors of Ref.~\cite{RMD-MM:10}
propose the formation of clusters through selective addition of {
  interconnections} to the network. More recently, the control of
partially synchronized states with applications to brain networks is
studied in Ref.~\cite{TM-GB-DSB-FP:19b} by means of structural
interventions, and in Ref.~\cite{KB-JOG-SHT-TV-JMV-SFM:19} via
exogenous stimulation. {
  Ultimately, the results presented in this work not only complement but go well beyond the
control of macroscopic synchronization observables and partial synchronization by} allowing to specify the synchrony
level of pairwise interactions.

{ While our contributions include the analysis of the stability
  properties of functional patterns, here we do not assess their
  basins of attraction. We emphasize that, in general, the estimation
  of the basin of attraction of the attractors of nonlinear systems
  remains an outstanding problem, and even the most recent results
  rely on numerical approaches or heavy modeling assumptions
  \cite{SC-MF-MM-GJP-VMP:21}. Further, in the case of coupled Kuramoto
  oscillators, existing literature shows that the number of equilibria
  for the phase differences increases significantly with the
  cardinality of the network \cite{DM-NSD-FD-JDH:15}, making the study
  of basins of attraction extremely challenging. In the Supplementary Information, we extend previous work on  identical oscillators to networks with heterogeneous oscillators, and show that that functional patterns can be at most
  almost-globally stable in cluster-synchronized positive networks. Yet, a precise estimation of the basin
  of attraction for any arbitrary target pattern goes beyond the scope of this
  work and may require the derivation of \emph{ad-hoc} principles
  based on Lyapunov's stability theory \cite{LZ-DJH:20}.  }

The framework presented in this work has { other} limitations,
which can be addressed in follow up studies.  First, despite its
capabilities in modeling numerous oscillatory network systems, the
Kuramoto model cannot capture the amplitude of the oscillations,
making it most suitable for oscillator systems where most of the
information is conveyed by phase interaction as demonstrated in
Ref.~\cite{Corbetta2015} for resting brain activity. { To model
  brain activity during cognitively demanding tasks such as learning,
  higher-dimensional oscillators may be more suitable
  \cite{YQ-TM-DSB-FP:20}.}  Second, the use of phase-locked
trajectories is instrumental to the control and design of functional
patterns. Yet, it is not necessary. In fact, restricting the control
to phase-locked dynamics does not capture exotic dynamical regimes in
which only a subset of the oscillators is frequency-synchronized.
Third and finally, in some situations the network parameters are not
fully known. While being still an active area of research, network
identification of oscillator systems may be employed in such scenarios
\cite{MJP-MC-LL-CMT-BX:19}.
 
Directions of future research can be both of a theoretical and
practical nature. For instance, follow up studies can focus on the
derivation of a general condition for the stability of a feasible
functional pattern in positive networks. Further, a thorough
investigation of which network structures allow for multiple prescribed
equilibria may be particularly relevant in the context of memory
systems, where different patterns are associated to different memory
states. Specific practical applications may also require the inclusion
of sparsity constraints on the accessible structural parameters for
the implementation of the proposed control and design framework.

\section*{Methods}

\small 
\subsection*{Matrix form of equation~\eqref{eq: kuramoto} and phase-locked solutions}
For a given network of oscillators, we let the entries of the
(oriented) incidence matrix $B$ be defined component-wise after
choosing the orientation of each interconnection $(i,j)$. In
particular, $i$ points to $j$ if $i<j$, and $B_{k\ell}=-1$ if
oscillator $k$ is the source node of the interconnection $\ell$,
$B_{k\ell}=1$ if oscillator $k$ is the sink node of the
interconnection $\ell$, and $B_{k\ell}=0$ otherwise. The matrix form
of equation~\eqref{eq: kuramoto} can be written as
\begin{align*}
\dot{\bs   \theta} = &~ [\omega_1 ~ \cdots ~ \omega_n]^\transpose - B \begin{bmatrix}\ddots & & \\ & \sin(x_{ij}) & \\ & & \ddots \end{bmatrix}  \bs  \delta\\
=&~ [\omega_1 ~ \cdots ~ \omega_n]^\transpose - B \diag(\{\sin(x_{ij})\}_{(i,j)\in\mc E})  \bs \delta =  [\omega_1 ~ \cdots ~ \omega_n]^\transpose - B D ({\bs x}) \bs  \delta,
\end{align*}
where $D({\bs x})$ is the diagonal matrix of the sine functions in equation~\eqref{eq: kuramoto}.

When the oscillators evolve in a phase-locked configuration, the
oscillator frequencies become equal to each other and constant. In
particular, since $\1^\transpose B = 0$, we have
$\1^\transpose \dot { \bs \theta} = \1^\transpose k\1 = \1^\transpose
[\omega_1 ~ \cdots ~ \omega_n]^\transpose$, thus showing that, in any
phase locked trajectory, the oscillators frequency $k$ needs to equal
the mean natural frequency $\frac{1}{n}\sum_{i=1}^n \omega_i$.

\subsection*{Any feasible functional pattern has $n-1$ degrees of freedom}
The values of a functional pattern can be uniquely specified using a
set of $n-1$ correlation values.  To see this, let us define the
incremental variables $\bs x = M\theta$, where
$M\in\real^{|\mc E|\times n}$ is the matrix whose $k$-th row,
associated to $x_{ij}$, is all zeros except for $b_{ki} = -1$ and
$b_{kj} = 1$.  Consider the first $n-1$ rows of $M$, associated to
$x_{12}, x_{13}, \dots x_{1n}$, and notice that they are linearly
independent. Moreover, the row associated to $x_{ij}$, $i>1$, can be
obtained by subtracting the row associated to $x_{1i}$ to the row
associated to $x_{1j}$, implying that the rank of $M$ is $n-1$. Any
collection of $n-1$ linearly independent rows of $M$ defines a full
row-rank matrix $M_\text{min}$ (e.g., any $n-1$ rows corresponding to
the transpose incidence matrix of a spanning tree
\cite{Godsil2001}). We let
$\bs x_\text{min} = M_\text{min} \bs \theta$, where $\bs x_\text{min}$
is a smallest set of phase differences that can be used to quantify
the synchronization angles among all oscillators. Because
$\ker(M_\text{min}) = \1$, we can obtain the phases $\bs \theta$ from
$\bs x_\text{min}$ modulo rotation:
$\bs \theta = M_\text{min}^\dagger \bs x_\text{min} - c\1$, where
$M_\text{min}^\dagger$ denotes the Moore-Penrose pseudo-inverse of
$M_\text{min}$ and $c$ is some real number. Further, since
$\ker(M_\text{min}) = \ker(M)$, we can reconstruct all phase
differences $\bs x$ from $\bs x_\text{min}$:
 \begin{equation*}
   M M_\text{min}^\dagger \bs x_\text{min} = M (\bs \theta + c\1) = M\bs \theta + 0 = \bs x.
 \end{equation*}

 The above equation reveals that all the differences $\bs x$ are
 encoded in $\bs x_\text{min}$. That is, any $x_{ij}$ can be written
 as a linear combination of the elements in $\bs x_\mathrm{min}$.  For
 example, if $n=3$ and
 $\bs x_\text{min} = [ x_{12}~ x_{23}]^\transpose$, then $x_{13}$ is a
 linear combination of the differences in $\bs x_\text{min}$, i.e.,
 $\bs x = \left[\begin{smallmatrix}-1 & 1 & 0\\ 0 & -1 & 1\\ -1 & 0 &
     1\end{smallmatrix}\right] \left[\begin{smallmatrix}-1 & 1 & 0\\ 0
     & -1 & 1\end{smallmatrix}\right]^\dagger \bs x_\text{min}$, in
 which $x_{13} = x_{12}+x_{23}$.  Thus, because $n-1$ incremental
 variables define all the remaining ones, the entries of any
 functional pattern must have only $n-1$ degrees of freedom.

{
\subsection*{Existence of a strictly positive solution to
  Problem~\eqref{eq: problem 2}}

Rewrite the pattern assignment problem
$B D({\bs x}) \bs \delta = { \bs \omega}$ as
\begin{align*}
  B D({\bs x}) \bs \delta = B_{:,\mc H}
  D_{\mc H,\mc H}({\bs x})\bs  \delta_{\mc H}  +
  B_{:,\tilde{\mc H}} D_{\tilde{\mc H},\tilde{\mc H}}({\bs x})\bs
  \delta_{\tilde{\mc H}} = \bs \omega,
\end{align*}
where the subscripts $\mc H$ and $\tilde{\mc H}$ denote the entries
corresponding to the Hamiltonian path in conditions (ii.a)-(ii.b) and the remaining ones,
respectively.   Since
$\mathrm{Im} (\bar B_{:,\mc H}) = \mathrm{Im} ( B_{:,\mc H}
D_{:,\mc H} ({\bs x}) )= \mathrm{span}(\textbf{1})^\perp$,
$\mathrm{Im} (B_{:,\tilde{\mc H}} D_{:,\tilde{\mc H}}({\bs x}))\subseteq \mathrm{span}(\textbf{1})^\perp$, and
$\bs \omega \in \mathrm{span}(\textbf{1})^\perp$, for any vector
$\bs \delta_{\tilde{\mc H}}$, the following set of weights solves the
above equation:
\begin{align*}
  \bs  \delta_{\mc H} &= 
 \left(B_{:,\mc H} D_{\mc H,\mc H}({\bs x})\right)^\dagger \left( \bs \omega - B_{:,\tilde{\mc H}} D_{\tilde{\mc H},\tilde{\mc H}}({\bs x})\bs
                        \delta_{\tilde{\mc H}} \right) =
                         ( D_{\mc H,\mc H}({\bs x}) B_{:,\mc H}^{\transpose} B_{:,\mc H} D_{\mc H,\mc H}({\bs x}))^{-1} D_{\mc H,\mc H}({\bs x}) B_{:,\mc H}^{\transpose} \left( \bs \omega - B_{:,\tilde{\mc H}} D_{\tilde{\mc H},\tilde{\mc H}}({\bs x})\bs
                        \delta_{\tilde{\mc H}} \right)
\end{align*}
Because the matrix $D_{\mc H,\mc H}({\bs x}) B_{:,\mc H}^{\transpose} B_{:,\mc H} D_{\mc H,\mc H}({\bs x})$ is
an M-matrix, its inverse has nonnegative entries. Further, by
condition (ii.b), $D_{\mc H,\mc H}({\bs x})  B_{:,\mc H}^\transpose \bs \omega$ is strictly
positive. Then, the vector
$( D_{\mc H,\mc H}({\bs x}) B_{:,\mc H}^{\transpose} B_{:,\mc H} D_{\mc H,\mc H}({\bs x}))^{-1} D_{\mc H,\mc H}({\bs x}) B_{:,\mc H}^{\transpose}  \bs \omega$ is also strictly positive, and so is the
solution vector $\bs \delta_{\mc H}$ for any sufficiently small and
positive vector $\delta_{\tilde{\mc H}}$.
 }

\subsection*{Enforcing stability of functional patterns in networks with cooperative and competitive interactions} To ensure that the Jacobian matrix in equation~\eqref{eq: Jacobian} is the Laplacian of a positive network and, thus, stable, we solve the problem in equation~\eqref{eq: problem 2} with a slight modification of the constraints. Specifically, we post-multiply the matrix $B$ in equation~\eqref{eq: problem 2} as $B\Delta$, where $\Delta=\diag(\{\mathrm{sign}(\cos( x_{ij}))\}_{(i,j)\in\mc E})$ is a matrix that changes the sign of the columns of $B$ associated to negative weights in the cosine-scaled network:
$$
 B\Delta D({\bs x})  \bs   \delta = { \bs  \omega}
 $$
  Solving for positive interconnection weights the problem in equation~\eqref{eq: problem 2} under the above modified constraint yields a stable Jacobian in a network where the final couplings are $\Delta \bs \delta$.

\subsection*{Heuristic procedure to promote stability of functional patterns in positive networks}
{ We provide a heuristic procedure to promote the stability of functional patterns that include negative correlations in a network with nonnegative weights. Our procedure relies on the definition of Gerschgorin disks and the Gerschgorin Theorem.
\smallskip

\noindent\textbf{Definition of Gerschgorin disk}. Let
$M\in\mathbb{C}^{n\times n}$ be a complex matrix. The $i$-th
Gerschgorin disk is $\mc D_i = (M_{ii}, r_i)$, $i=1,\dots,n$, where the
radius is $r_i = \sum_{j\ne i} |M_{ij}|$ and the center is $M_{ii}$.
\smallskip

\setcounter{appxthm}{1}
\renewcommand{\theappxthm}{\arabic{appxthm}}

\begin{appxthm}{\emph{({\bfseries Gerschgorin}
      \cite{Gerschgorin1931})}}. \textit{The eigenvalues of the matrix
    $M$ lie within the union $\bigcup_{i=1}^n D_i$ of its Gerschgorin
    disks.}
\end{appxthm}

Whenever all target phase differences in ${\bs x}$ satisfy
$| x_{ij}|\le\frac{\pi}{2}$, the Gerschgorin disks of the Jacobian
in equation~\eqref{eq: Jacobian} all lie in the closed left
half-plane. However, for patterns ${\bs x}$
containing phase differences $| x_{ij}|\ge\frac{\pi}{2}$, the
union of the Gerschgorin disks intersects the right
half-plane. Reducing the magnitude of the entries satisfying
$A_{ij}\cos( x_{ij})<0$ effectively shrinks the radius of the
Gerschgorin disks that overlap with the right half-plane and shifts their centers towards the left-half plane due to the structure of the Jacobian matrix. We remark
that the procedure in equation~\eqref{eq: heuristic} is a heuristic,
and it is provably effective only when all interconnections with
$A_{ij}\cos( x_{ij})<0$ can be removed, so that all the Gerschgorin disks lie completely in the left-half plane.
}

\subsection*{Data availability}
{ The brain data that support the findings of this study are available from the Supplementary Information of \cite{Corbetta2015}, \url{https://doi.org/10.1371/journal.pcbi.1004100.s006}.
The IEEE 39 New England data parameters and interconnection scheme analyzed in this study for the structure-preserving power network model can be found in the reference textbook \cite{PWS-MAP:98} (see also Supplementary Information for modeling assumptions). The parameters for the simulations on the network-reduced IEEE 39 New England test case in the Supplementary Information have been obtained from Ref.~\cite{YS-IM-TH:11} and Ref.~\cite{AM-IK-PB-GS:15}.}

\subsection*{Code availability}
{ Source code and documentation for the numerical simulations presented here are freely available in GitHub at: \url{https://github.com/tommasomenara/functional_control} with the identifier \texttt{10.5281/zenodo.4546413}.}

\normalsize

\bibliographystyle{plain}

\bibliography{alias,FP,Main,New}

\newpage
\section*{Supplementary Information}

\section{Supplementary Text}

 \subsection{Comparison between our correlation metric and the Pearson correlation coefficient}

\begin{figure}[h]
\centering
\includegraphics[width=.6\textwidth]{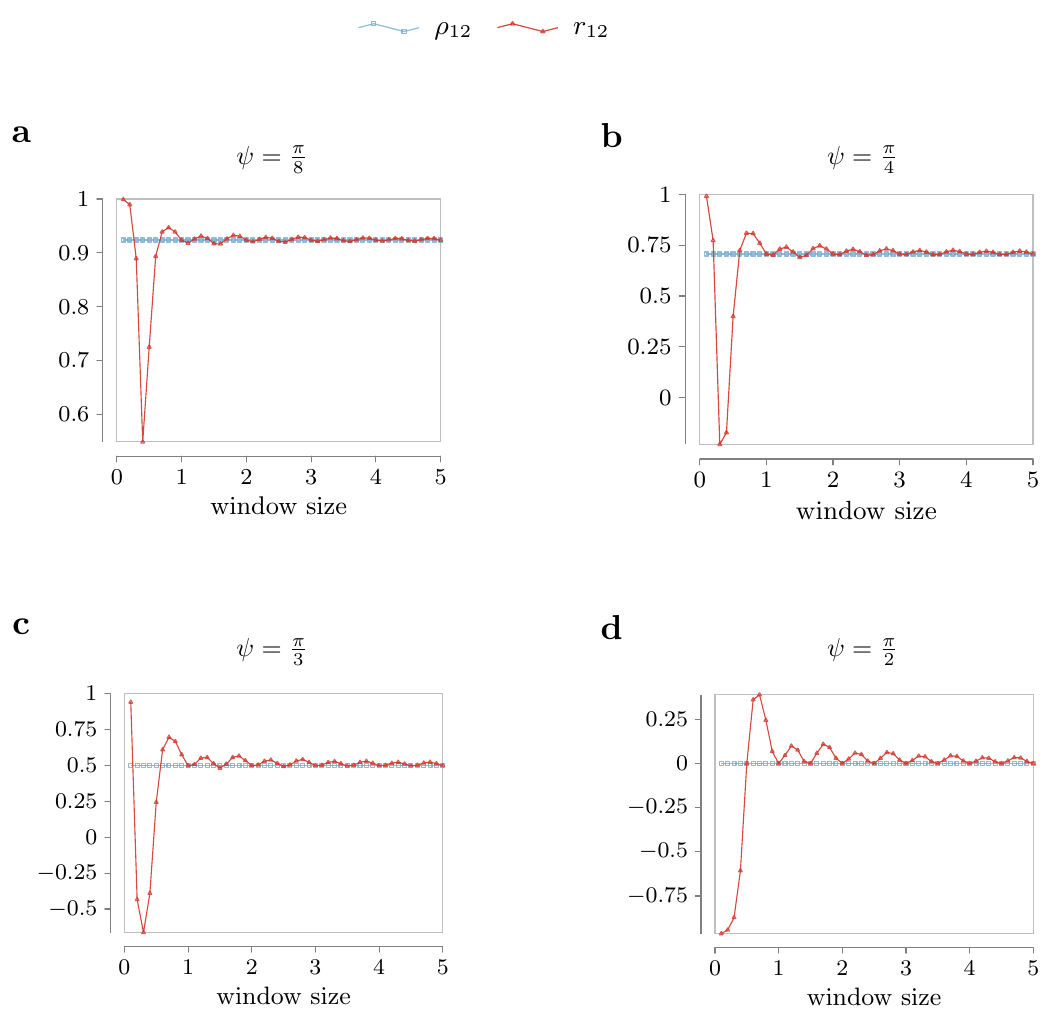}
 \vspace{0.4cm}
\caption{{ {\bf Comparison between Pearson correlation coefficient and the metric $\rho_{12} = <\cos(\theta_2-\theta_1)>_{t}$ on two phase-locked signals and varying time window lengths.} Each point in the plot represents a value of $\rho_{12}$ (in blue) and $r_{12}$ (in red) computed in a time window $[0~T]$, where $T=0.1,0.2,\dots,5$. It can be seen in all panels that $\rho_{12}$ is only affected by the phase shift $\psi$. Instead, the Pearson correlation coefficient returns oscillating values for different window sizes with damping oscillations as the length of the time window increases. {\bf a} The phase shift in the initial conditions $\psi = \frac{\pi}{8}$. {\bf b} The phase shift in the initial conditions $\psi = \frac{\pi}{4}$. {\bf c} The phase shift in the initial conditions $\psi = \frac{\pi}{3}$. {\bf d} The phase shift in the initial conditions is $\psi = \frac{\pi}{2}$.} \label{fig: Pearson vs cos}}
\end{figure}

{ In this work, we provide methods and principles to guarantee the emergence of phase-locked trajectories that describe functional patterns. The latter are defined utilizing $\rho_{ij} = <\cos(\theta_j-\theta_i)>_{t}$, instead of the classical Pearson correlation coefficient. The same metric is also utilized in \cite{GD-JC-JC-ET-HL-NKL-MLK:19} to quantify the similarity between BOLD signals in functional MRI recordings, and in \cite{AA-AD-CJP:06}, to quantify the synchrony in hierarchical networks.

To compare the two metrics, recall the definition of the classical (sample) Pearson correlation, which is a measure of linear correlation between two series of data,  for vectors of samples $y\in\real^T$ and $z\in\real^T$:
$$
r_{ij} = \frac{\sum_{i=1}^N (y_i- y_\mathrm{mean}) (z_i- z_\mathrm{mean})}{\sqrt{\sum_{i=1}^N (y_i- y_\mathrm{mean})^2} \sqrt{\sum_{i=1}^N (z_i- z_\mathrm{mean})^2}},
$$
where $y_\mathrm{mean}$ and $z_\mathrm{mean}$ denote the sample means of vectors $y$ and $z$, respectively.
Notice that the length $T$ of the vectors $y$ and $z$ depends on the sampling time and on the window length. We argue that our metric (equation~(2) in the main text) is simply more convenient when dealing with periodic phase signals. While we could use the Pearson correlation coefficient to define functional pattern instead of our cosine-based metric, the former does not perform well on periodic signals that evolve on the unit circle, and is heavily dependent on the time window employed to collect the samples. Thus, specific adjustments are needed to define correlation patterns through Pearson correlation coefficient for the class of time series (phase trajectories) studied in this work.

As an example, consider two identical sinusoidal signals (i.e., phase-locked) with natural frequency $\omega = 2\pi$ but shifted initial conditions: $\theta_1(0) = 0$, $\theta_2(0) = \varphi$, with $\varphi \in (0,~\pi]$. Fig.~\ref{fig: Pearson vs cos} illustrates the differences between the Pearson correlation coefficient $r_{12}$ and our correlation metric $\rho_{12}$ computed over a time window of varying length and for different values of the initial phase shift $\varphi$. In all panels, the values of $r_{12}$ vary at each point, emphasizing the dependence of the Pearson correlation coefficient from the length of the time window. Conversely, in all panels, the value of $\rho_{12}$ remains unaltered by the choice of time window length.

In conclusion, we choose to define functional patterns through $\rho_{ij} = <\cos(\theta_j-\theta_i)>_{t}$ because it is a convenient and suitable metric for this class of phase trajectories that naturally emerge in oscillator systems.
}

\subsection{Stability results for functional patterns with angle differences larger than $\frac{\pi}{2}$ in the case of line and cycle with positive weights}
Consider a line network of $n$ (ordered) oscillators with positive-only weights that possesses an equilibrium for the phase difference dynamics satisfying $|x_{i,i+1}|>\frac{\pi}{2}$ for some $i\in\{1,\dots,n-1\}$. It is straightforward to deduce from the results on structural balance \cite{DZ-MB:17} that the considered equilibrium is unstable. This implies that the functional pattern associated with that equilibrium is unstable.

Intuitively, the simplest topology that lends itself to a characterization of stable phase configurations including $|x_{i,j}|>\frac{\pi}{2}$ and allowing only positive weights is the cycle (i.e., a line network where the first and last oscillators are connected). Consider a cycle network of $n>4$ oscillators with positive weights{, and denote with ${\bs x}_\mathrm{cycle}$ a vector of the $n$ phase differences between connected oscillators}. We find that, after a suitable relabeling of the oscillators { for which $x_{12}$ satisfies $|x_{12}|>\frac{\pi}{2}$}:

\setcounter{appxthm}{1}
\renewcommand{\theappxthm}{\arabic{appxthm}}
\begin{appxthm}{({\bfseries Stability of phase differences equilibria with $|x_{ij}|>\frac{\pi}{2}$ in cycle networks})}\label{thm: cycle}
The equilibrium ${\bs x}_\mathrm{cycle} = [x_{12}~ x_{23}~ \dots~ x_{n-1,n}]^\transpose = [\gamma~ \varphi_1~ \dots~ \varphi_{n-1}]^\transpose$ with $|\gamma|>\frac{\pi}{2}$, $|\varphi_i|<\frac{\pi}{2}$, and $\varphi_{i} = \varphi_{n-i}$ for all $i=1,\dots,n-1$, is stable if and only if
\begin{itemize}
\item[\emph{(i)}] $A_{i,i+1} = A_{12}\frac{\sin(\gamma)}{\sin(\varphi_{i-1})}$\\[.3em] for $i = 2,\dots,n$, with $n-i+3 \triangleq 1$ if $i = 2$, and ${ n+1} \triangleq 1$;
\item[\emph{(ii)}] $|\cot{\gamma}| \le \left(\tan(\varphi_1)+\dots+\tan(\varphi_{n-1})\right)^{-1}$.
\end{itemize}
Moreover, if $\varphi_1 = \dots = \varphi_{n-1}$ and $n\to\infty$, the largest possible value for $\gamma$ { such that $\bs x_\mathrm{cycle}$ is stable} tends to the value $\gamma \approx 1.789776$, { which is the solution to $\gamma-\tan(\gamma)=2\pi$.}
\end{appxthm}
\smallskip

\begin{pfof}{Theorem \ref{thm: cycle}} To assess the stability of the equilibrium $\bar{\bs x}_\mathrm{cycle} = [x_{12}~ x_{23}~ \dots~ x_{n-1,n}]^\transpose = [\gamma~ \varphi_1~ \dots~ \varphi_{n-1}]^\transpose$ with $|\gamma|>\frac{\pi}{2}$, $|\varphi_i|<\frac{\pi}{2}$, and $\varphi_{i} = \varphi_{n-i}$ for all $i=1,\dots,n-1$, we analyze the spectrum of the Jacobian $J({\bs x}_\mathrm{cycle}) = -\mc L({\bs x}_\mathrm{cycle})$. From Ref.~\cite[Corollary IV.7]{DZ-MB:17}, we have that a necessary and sufficient condition for the Laplacian matrix $\mc L({\bs x}_\mathrm{cycle})$ of the cosine-scaled network to be positive semidefinite is
\begin{equation}\label{eq: resistance condition}
|{ A_{12}}\cos(\gamma)| \le \mc R_{12}^{-1},
\end{equation}
with $\mc R_{12}$ being the effective resistance of the graph in which the edge $(1,2)$ has been removed. That is,
\begin{equation}\label{eq: resistance}
\mc R_{12} = \frac{1}{{ A_{23}}\cos(\varphi_1)} + \dots + \frac{1}{{ A_{n-1,n}}\cos(\varphi_{n-1})}.
\end{equation}
Since the adjacency matrix satisfies $A=A^\transpose$ and ${\bs x}_\mathrm{cycle}$ is an equilibrium for the difference dynamics of the cycle network, the network weights must be identical pairwise:
\begin{equation}\label{eq: symmetric weights}
A_{i,i+1} = A_{n-i+2,n-i+3} = A_{12}\frac{\sin(\gamma)}{\sin(\varphi_{i-1})},
\end{equation}
for $i = 2,\dots,n$ with the convention $n-i+3 \triangleq 1$ if $i = 2$. Thus, the angles being identical pairwise implies that the network weights must also be identical pairwise, which yields condition (i) of Theorem \ref{thm: cycle}. Moreover, plugging the network weights from equation~\eqref{eq: symmetric weights} into { equation~\eqref{eq: resistance}} yields
\begin{equation}
\mc R_{12} = \frac{\tan(\varphi_1)}{{ A_{12}}\sin(\gamma)} + \dots + \frac{\tan(\varphi_{n-1})}{{ A_{12}}\sin(\gamma)},
\end{equation}
which makes the condition in   Eq.~\eqref{eq: resistance condition} become, after { algebraic} calculations, condition (ii) of Theorem \ref{thm: cycle}:
\begin{equation}\label{eq: cycle condition}
|\cot{\gamma}| \le \left(\tan(\varphi_1)+\dots+\tan(\varphi_{n-1})\right)^{-1}.
\end{equation}
Thus, given that the Laplacian $\mc L({\bs x}_\mathrm{cycle})$ is positive definite, the Jacobian $J({\bs x}_\mathrm{cycle})$ is stable, and its only zero eigenvalue is due to rotational symmetry of the right-hand side of the Kuramoto dynamics {(equation~(2) of the main manuscript)}.

For $\varphi_1 = \dots = \varphi_{n-1}$, we have that
$$
\varphi_i = \frac{2\pi-\gamma}{n-1}.
$$
Hence, the right-hand side of { equation}~\eqref{eq: cycle condition} becomes $\frac{\cot(\varphi)}{n-1}$, and $\lim_{n\to\infty}\frac{\cot(\varphi)}{n-1} = \frac{1}{2\pi-\gamma}$. Since $|\gamma|>\frac{\pi}{2}$, plugging the limit value for $\frac{1}{2\pi-\gamma}$ into   { equation}~\eqref{eq: cycle condition} and solving for the equality yields $\gamma - \tan(\gamma) = 2\pi$, whose unique solution is $\gamma \approx 1.789776$. This concludes the proof.
\end{pfof}

\subsection{Functional patterns possess at most almost-global stability in positive cluster-synchronized networks}
{
In general, the analysis and estimation of the basin of attraction of nonlinear systems remains an outstanding problem, and even the most recent results rely on numerical approaches or heavy modeling assumptions \cite{SC-MF-MM-GJP-VMP:21}. To the best of our knowledge, Ref.~\cite{LZ-DJH:20} provides the most up-to-date study of the basins of attraction of synchronized Kuramoto oscillators. However, the authors in Ref.~\cite{LZ-DJH:20} only derive estimates of the basin of attraction for the fully synchronized case in networks of identical oscillators. Below, we show that functional patterns associated with phase-locked trajectories in cluster-synchronized positive networks are at most almost-globally stable. Our results extend previous work on identical oscillators to the case of oscillators with heterogeneous natural frequencies.

Existing literature shows that the number of equilibria for the phase differences of heterogeneous oscillators evolving on random graphs increases significantly with the cardinality of the network \cite{DM-NSD-FD-JDH:15}. Therefore, we begin by restricting our analysis to the case of identical oscillators. In general, for any connected network of identical oscillators with cooperative connections $A_{ij}\ge 0$, there is no unique pattern. In fact, the largest basin of attraction is achieved by $\mathbb{S}^1$\emph{-synchronizing} graphs (e.g., complete graphs, acyclic graphs, and sufficiently dense graphs \cite{Doerfler2014,AT-MS-SHS:20}), which feature almost-global stability of the fully synchronized functional pattern. In this class of networks, the only stable equilibrium ${\bs x}$ satisfies $ x_{ij} = 0$ for all $i,j$, and there exists a finite number of unstable equilibria. For instance, consider two connected identical oscillators: $\omega_1 = \omega_2$. The only stable equilibrium for the phase differences dynamics is $x_{12} = \theta_2-\theta_1 = 0$. Yet, there also exists an unstable equilibrium $x_{12} = \pi$. Note that, in general, almost-global stability cannot be guaranteed for arbitrary classes of sparse networks of identical oscillators, as other synchronization manifolds besides the fully synchronized one can be stable -- for example, splay states emerge in Cayley graphs \cite{GSM-XT:15}. This latter class of graphs admits two stable equilibria (full synchronization and splay states), which implies the coexistence of two stable patterns.

We can generalize the above observations to networks of heterogeneous oscillators. To do so, we leverage cluster synchronization -- a phenomenon where distinct groups of synchronized oscillators coexist in a network \cite{TM-GB-DSB-FP:18} -- of $\mathbb{S}^1$\emph{-synchronizing} clusters with possibly different natural frequencies. We consider cluster synchronization in networks with a partition of the oscillators $\mc C = \{\mc C_1, \dots, \mc C_m \}$, with $\mc C_k \subseteq \mc O$ being a subset of the network oscillators constituting an $\mathbb{S}^1$\emph{-synchronizing} graph, $k\in\until{m}$, where $\bigcup_{k=1}^m \mc C_{k} = \mc O$ and $\mc C_k \cap \mc C_{\ell} = \emptyset$ if $k \ne \ell$. Further, $\omega_i = \omega_j$ for every $i,j\in\mc C_k$, $k\in\until{m}$.\footnote{This is a necessary and sufficient condition for cluster synchronization \cite{TM-GB-DSB-FP:18}.} Whenever cluster synchronization emerges, the diagonal blocks (of sizes $|\mc C_k|\times |\mc C_k|$, $k=1,\dots,m$) of the functional pattern associated with partition $\mc C$ satisfy $\rho_{ij} = 1$ (see Supplementary Fig.~\ref{fig: cluster synch}).\footnote{Without loss of generality, we assume that the network oscillators are labeled such that consecutive labels belong to the same cluster.} Thus, for a given number of clusters $m\ge 1$, each of the ${n \choose m}$ possible ways to partition the network yields a functional pattern with diagonal blocks corresponding to synchronized clusters.

By extending the above observations on the stability of $\mathbb{S}^1$\emph{-synchronizing} graphs, we find that, in networks where at least one cluster satisfies $|\mc C_k|\ge 2$, at most almost-global stability of cluster-synchronized functional patterns can be achieved. In fact, for any choice of intra-cluster phase differences from the finite set of unstable equilibria in $\mathbb{S}^1$\emph{-synchronizing} graphs (i.e., $x_{ij}=\pi$ for at least one intra-cluster phase difference), there exists a value for the inter-cluster phase differences for which the network admits phase-locked trajectories where intra-cluster phase differences satisfy $x_{ij} = 0$ or $x_{ij} = \pi$.

As an example, consider a $4$-oscillator network partitioned as $\mc C = \{\mc C_1, \mc C_2\}$, where $\mc C_1 = \{1,2\}$ and $\mc C_2 = \{3,4\}$. The network parameters read
$$
A = \begin{bmatrix} 0& 5 & 2 & 0\\5 & 0 & 0 & 2\\ 2 & 0 & 0 & 6\\ 0 & 2 & 6 & 0 \end{bmatrix} \ \text{ and } \ \bs\omega = \begin{bmatrix} 1 \\ 1\\2\\2\\ \end{bmatrix}.
$$
Note that the two clusters are $\mathbb{S}^1$\emph{-synchronizing} graphs.
It can be shown that there exists an unstable equilibrium at ${\bs x}_\mathrm{desired} = [ x_{12}~ x_{23}~ x_{34}]^\transpose = [\pi~0.25268~\pi]^\transpose$. Thus, whenever the pattern $R$ associated to the partition $\mc C$ is stable, it is at most almost-globally stable.
}

\subsection{Allocation of multiple phase-locked equilibria by tailoring of the network structural parameters}\label{sec: multi equilibria}

The convex optimization problem proposed in the main text allows to tailor the network weights and the natural frequencies of the oscillators to specify multiple equilibria for the dynamics of the phase differences ${\bs x}$. These equilibria correspond to phase trajectories $\theta$ that evolve with constant, desired phase differences.

We now provide an example where we jointly impose, for a complete graph of $n=7$ oscillators, two equilibria for the phase difference dynamics. Specifically, by taking $\theta_1$ as a reference, we choose two points for the phase differences $x_{1i} = \theta_i-\theta_1$ to be set as equilibria: $ {\bs x}^{(1)}_\mathrm{desired} = \left[\frac{\pi}{6}~ \frac{\pi}{6}~ \frac{\pi}{4}~ \frac{\pi}{4}~ \frac{\pi}{6}~ \frac{\pi}{4}\right]^\transpose$ and $ {\bs x}^{(2)}_\mathrm{desired} = \left[\frac{\pi}{8}~ \frac{\pi}{3}~ \frac{\pi}{4}~ \frac{\pi}{4}~ \frac{\pi}{6}~\frac{\pi}{4}\right]^\transpose$. The initial network parameters (adjacency matrix and zero-mean natural frequencies) read as:
$$
A = \begin{bmatrix}
0 & 2 & 2 & 2 & 2 & 2 & 2 \\
    2  &       0 & 2  &  2  &  2  &  2  &  2\\
    2 &  2    &     0  &  2  &  2  &  2  &  2\\
  2 &   2 &   2   &      0  &  2  &  2 &   2\\
   2  &  2  &  2  &  2     &    0  &  2  &  2\\
    2  &  2  &  2  &  2  &  2    &     0  &  2\\
   2  &  2  &  2  &  2  &  2 &   2   &      0
\end{bmatrix} \ \text{and} \ 
{\bs \omega} = \begin{bmatrix}0.3160\\
   -0.1266\\
    0.1437\\
    0.2771\\
   -0.3593\\
   -0.3363\\
    0.0854\end{bmatrix},
$$
respectively. To impose the desired equilibria, we numerically solve the following convex problem through standard \texttt{cvx} routines \cite{MG-BS-YY:09}:
\begin{align*}
\min_{\bs \alpha} &\ \ \ \|\bs\delta+\bs\alpha \|_1\\
\text{subject to}  &\ \ \begin{bmatrix}B D\left({ \bs  x}^{(1)}\right)\\[.5em] B D\left({\bs x}^{(2)}\right) \end{bmatrix} (\bs\delta+\bs\alpha) = \begin{bmatrix}{ \bs \omega}\\ { \bs \omega}\end{bmatrix}.
\end{align*}
The solution $\bs\alpha^*$ yields the following corrected adjacency matrix:
$$
 A_\mathrm{c} = \begin{bmatrix}
      0  &  4.2841  &  1.3731 &  -1.6720 &  -2.5576  &  2.4333 &  -1.9382\\
    4.2841  &       0 &  -3.0379  &  2.9361  &  2.9041  &  2  &  2.9252\\
    1.3731 &  -3.0379    &     0  &  0.7026  &  0.6949  &  0.5221  &  0.7\\
   -1.6720 &   2.9361 &   0.7026   &      0  &  2  &  2 &   2\\
   -2.5576  &  2.9041  &  0.6949  &  2     &    0  &  2  &  2\\
    2.4333  &  2  &  0.5221  &  2  &  2    &     0  &  2\\
   -1.9382  &  2.9252  &  0.7  &  2  &  2 &   2   &      0
\end{bmatrix}.
$$
An investigation of the Jacobian spectrum (see main text for results on stability) of the phase differences computed at the two equilibria $ {\bs x}^{(1)}_\mathrm{desired}$ and $ {\bs x}^{(2)}_\mathrm{desired}$ reveals that the first equilibrium point is unstable and that the second one is locally stable.
We illustrate the outcome of the above procedure to specify multiple equilibria in Fig.~6{\bf b} in the main text, where the phase differences start at $ {\bs x}^{(1)}_\mathrm{desired}$ at time $t=0$, and converge to $ {\bs x}^{(2)}_\mathrm{desired}$ after a perturbation is applied at $t=50$ to force them out from the equilibrium $ {\bs x}^{(1)}_\mathrm{desired}$.

\subsection{A heuristic method to promote stability of functional patterns in positive networks}

To promote stability of functional patterns with phase differences $|x_{ij}|>\frac{\pi}{2}$, it is advantageous to design the network weights by minimizing the ones associated to a $\cos(x_{ij})<0$ (i.e., reducing $A_{ij}$ as much as possible) in the cosine-scaled network. Reducing the magnitude of these coupling strengths, or even pruning such interconnections, causes the Gerschgorin disks of the Laplacian $\mc L({\bs x}_\mathrm{desired})$ to lie almost entirely in the right half-plane. In fact, by considering the limit case where all negative connections in the cosine-scaled network are pruned, if the remaining connections describe a connected network, then the Laplacian becomes a classical positive semi-definite Laplacian. This guarantees stability of the desired functional pattern and motivates the following heuristic procedure.

To showcase the effectiveness of the proposed heuristic method to promote stability, we construct an example with $n=7$ oscillators, { $|\mc E| = 9$ interconnections}, and desired minimum vector of phase differences
$$
 {\bs x}_\text{desired}^{(1)} = \left[\frac{ 21\pi}{32} ~ \frac{\pi}{6} ~\frac{\pi}{6} ~\frac{\pi}{8} ~\frac{\pi}{8} ~\frac{\pi}{3}\right]^\transpose,
$$
where $x_{ij} = \theta_j-\theta_1$, $j = 2,\dots, 7$. Notice that the first difference $x_{12} > \pi/2$, hence $\cos(x_{12})<0$.
Consider the oscillator network with structural parameters that read as:
\begin{equation}\label{eq: initial}
A = \begin{bmatrix}
            0  &  0.1706   &      0  &       0  &  0.5796   &      0    &     0\\
    0.1706     &    0  &  1.3434  &       0   &      0   &      0   &      0\\
         0  &  1.3434     &    0  &  2  &  2 &    2.2140    &     0\\
         0    &     0 &   2    &     0    &     0  &  1.2392 &   0.3432\\
    0.5796    &     0  &  2 &        0   &      0  &  2 &        0\\
         0   &     0  &  2.2140  &  1.2392 &   2    &     0     &    0\\
         0   &      0       &  0  &  0.3432    &     0     &    0      &   0\\
\end{bmatrix} \ \text{and} \ 
\bar{\bs \omega} = \begin{bmatrix}-0.3424\\
    0.4683\\
    0.2023\\
   -0.0099\\
   -0.0393\\
   -0.4507\\
    0.1716\end{bmatrix}.
\end{equation}
Such a network admits the following phase-locked equilibrium
$$
 {\bs x}_\text{desired}^{(0)} = \left[\frac{\pi}{4} ~ \frac{\pi}{6} ~\frac{\pi}{6} ~\frac{\pi}{8} ~\frac{\pi}{8} ~\frac{\pi}{3}\right]^\transpose,
$$
 which differs from $ {\bs x}_\text{desired}^{(1)}$ only in $x_{12}$, and has all $x_{ij}$ satisfying $|x_{ij}|<\pi/2$ (thus generating a stable functional pattern). Supplementary Fig.~\ref{fig: heuristic}\textbf{a}-\textbf{b} illustrate the stability of $ {\bs x}_\text{desired}^{(0)}$ and the functional pattern $R_0$ associated with this equilibrium.

The network considered in this example has $9$ interconnections that can be modified, and the desired equilibrium ${\bs x}_\text{desired}^{(1)}$ implies that $\mc N = \{1 \}$, { so that $\bs\alpha_{\mc N}$ is the modification of the first interconnection. To compute the optimal tuning of the network weights we solve the optimization (through standard \texttt{cvx} routines \cite{MG-BS-YY:09})
\begin{align}\label{eq: heuristic}
\min_{\bs  \alpha} &\ \ \left\| \bs\delta_{\mathcal N} + \bs \alpha_{\mathcal N} \right\|_{1}\\
\mathrm{subject~to }&\ \ B D({\bs x}) (\bs \delta + \bs\alpha) = {\bs  \omega}, \notag\\
\mathrm{and}&\ \ (\bs \delta + \bs \alpha) \ge 0 \notag.
\end{align}
 The optimal $\bs\alpha^*$ reads
$$\bs\alpha^* = [
   -0.1706~
    0.3152~
   -0.8748~
    59~
    0.9242~
   0~0~0~59]^\transpose,
$$
and the adjusted network adjacency matrix becomes:
\begin{equation}\label{eq: adjusted}
\tilde A = \begin{bmatrix}
                  0  &  {\color{red}\boldsymbol 0}    &    0    &     0   & 0.8948   &      0     &    0\\
    {\color{red}\boldsymbol 0}     &     0   & { 0.4686}   &      0 &        0   &      0   &   0\\
         0 &  { 0.4686}   &      0 &   61 &   2.9242  &  2.2140     &    0\\
         0  &       0   & 61 &        0    &     0  &  1.2392  &  0.3432\\
    0.8948     &    0 &   2.9242   &      0    &    0  &  61    &     0\\
         0   &      0    & 2.2140 &   1.2392 &   61   &      0    &     0 \\
         0   &      0     &    0   & 0.3432    &     0   &      0      &   0\\
\end{bmatrix},
\end{equation}
}where the optimal network correction $\bs\alpha^*$ has disconnected the entry $A_{12}$ (highlighted in red). The matrix $\tilde A$ guarantees stability of the functional pattern associated with $ {\bs x}_\text{desired}^{(1)}$, as we illustrate in Supplementary Fig.~\ref{fig: heuristic}\textbf{c}-\textbf{d}.

{
In practice, one may jointly optimize the network weights to satisfy the equilibrium constraints and the heuristic stability strategy while trying to keep the overall modification as small as possible. To do so, we let $\mathcal P$ (resp., $\mathcal N$) denote the set of indices associated with $A_{ij}\cos(x_{ij})>0$ (resp., $A_{ij}\cos( x_{ij})<0$). Then, the optimization problem that enacts the proposed strategy reads as:
\begin{align}\label{eq: heuristic2}
\min_{\bs  \alpha} &\ \  c_1\left\| \bs\alpha_{\mathcal P} \right\|_{\star} + c_2\left\| \bs\delta_{\mathcal N} + \bs \alpha_{\mathcal N} \right\|_{\star}\\
\mathrm{subject~to }&\ \ B D({\bs x}) (\bs \delta + \bs\alpha) = {\bs  \omega}, \notag\\
&\ \ (\bs \delta + \bs \alpha) \ge 0 \notag,
\end{align}
where $\| \cdot \|_\star$ is a desired vector norm, $c_1,c_2>0$ are arbitrary penalty { coefficients}, $\bs \alpha_{\mc P}$ denotes the entries of the tuning vector $\bs \alpha$ that are associated to positive weights in the cosine-scaled network, and $\bs \alpha_{\mc N}$ denotes the entries of the tuning vector $\bs \alpha$ that are associated to negative weights $\bs \delta_{\mc N}$ in the cosine-scaled network.

To compute the optimal tuning of the network weights in the same $7$-oscillator network above, we solve the optimization in Eq.~\eqref{eq: heuristic2} with $c_1 = 0.1$ and $c_2 = 10$ by minimizing the $\ell_1$-norm. The optimal $\bs\alpha^*$ reads
$$\bs\alpha^* = [
   -0.1706~
    0.3152~
   -0.8748~
    0~
    0.9242~
   0~0~0~0]^\transpose,
$$
and the adjusted network adjacency matrix becomes:
\begin{equation*}
\tilde A = \begin{bmatrix}
                  0  &  {\color{red}\boldsymbol 0}    &    0    &     0   & 0.8948   &      0     &    0\\
    {\color{red}\boldsymbol 0}     &     0   &  0.4686   &      0 &        0   &      0   &   0\\
         0 & 0.4686   &      0 &   2 &   2.9242  &  2.2140     &    0\\
         0  &       0   & 2 &        0    &     0  &  1.2392  &  0.3432\\
    0.8948     &    0 &   2.9242   &      0    &    0  &  2    &     0\\
         0   &      0    & 2.2140 &   1.2392 &   2   &      0    &     0 \\
         0   &      0     &    0   & 0.3432    &     0   &      0      &   0\\
\end{bmatrix},
\end{equation*}
where, as in the procedure above, the optimal network correction $\alpha$ has disconnected the entry $A_{12}$ (highlighted in red). Supplementary Fig.~\ref{fig: eig
  shift2} illustrates the shift of the Jacobian's eigenvalues while the
optimal tuning vector $\bs\alpha^*$ is gradually applied. The main differences with respect to the minimization in \eqref{eq: heuristic} is that the norm of $\bs\alpha^*$ is smaller, and the eigenvalues of $\tilde A$ are closer to the original ones in $A$.
}

\subsection{Extension of the proposed optimization methods to directed networks}\label{sec: asymmetry} 
{
By relaxing the assumption on symmetric adjacency matrices, the matrix form of an oscillator network with Kuramoto dynamics in equation~(3) of the main text does not hold anymore and requires a rewriting. In what follows, we use the subscript ``$\mathrm{d}$" to indicate notation associated to \emph{directed} graphs. Specifically, $\mc E_\mathrm{d}$ denote the oriented edge set, so that $D_\mathrm{d}\in\real^{|\mc E_\mathrm{d}| \times \mc |\mc E_\mathrm{d}|}$ and $\bs\delta_\mathrm{d}\in\real^{|\mc E_\mathrm{d}|}$ denote the diagonal matrix of all $\sin(x_{ij})$ with $(j,i)\in\mc E_\mathrm{d}$, and the vector of all the network weights $A(i,j)\ne 0$, respectively. Further, let us define $B_\mathrm{source}\in\real^{n\times|\mc E_\mathrm{d}|}$ the modified incidence matrix whose columns have nonzero entries only at the edges' sources. That is, $B_{\mathrm{source},{k\ell}}=-1$ if $k$ is the source of the interconnection $\ell$, and $B_{\mathrm{source},{k\ell}}=0$ otherwise. These definitions allow us to define the matrix form for a directed network of Kuramoto oscillators:
\begin{equation}\label{eq: directed matrix form}
\dot{\bs\theta} = [\omega_1~\cdots~\omega_n]^\transpose - B_\mathrm{source}D_\mathrm{d}(\bs x)\bs\delta_\mathrm{d}.
\end{equation}

The main change in the behavior of directed networks with respect to undirected ones is that the frequencies $\dot{\bs\theta}$ of phase-locked trajectories do not typically converge to the average natural frequency (i.e., $\dot{\bs\theta} \ne \omega_\mathrm{mean}\1$). For phase-locked trajectories we have that $\dot{\bs\theta} =\omega_\mathrm{sync}\1$, where the constant $\omega_\mathrm{sync}\in\real$ is not known \emph{a priori}, and can only be estimated in the almost-fully synchronized regime $|\theta_i(t)-\theta_j(t)|\ll 1$ for all $i,j\in\mc O$ \cite{PSK-DT-JS-AA:16}. Yet, not knowing $\omega_\mathrm{sync}$ as we do in the undirected case does not prevent the definition of a framework that can enforce a target functional pattern. In fact, by using equation~\eqref{eq: directed matrix form} we can concurrently achieve the functional pattern associated with $\bar{\bs x}_\mathrm{desired}$ and assign a desired phase-locked frequency $\omega_\mathrm{sync}$. To do so, we utilize
equation~\eqref{eq: directed matrix form} above with $\dot{\bs\theta} = \omega_\mathrm{sync}\1$ as a constraint in the optimization problems proposed in the main text. This modification allows us to extend any of the control methods to achieve desired functional relationships to directed networks.

As an example of optimization of the coupling strengths, we solve
\begin{align}\label{eq: optimization 3}
\min_{\bs \alpha} &&&  \left\| \bs \alpha \right\|_{1} \\
\mathrm{subject~to }&&& \bs\omega - B_\mathrm{source}D_\mathrm{d}(\bs x)(\bs\delta_\mathrm{d}+\bs \alpha) = \omega_\mathrm{sync}\1, \tag{\theequation a}\label{eq: constraint 2}
\end{align}
for a network of $n=7$ oscillators with adjacency matrix and natural frequencies as follows:
$$
A_\mathrm{d} = \begin{bmatrix}
0 & {\rd \bs 0} & 1 & 1 & 1 & 1 & 1 \\
    1  &       0 & 1  &  1  &  1  &  1  &  1\\
    1 &  1    &     0  &  1  &  1  &  1  &  1\\
  1 &   1 &   1   &      0  &  1  &  1 &   1\\
   1  &  1  &  1  &  1     &    0  &  1  &  1\\
    1  &  1  &  1  &  1  &  1    &     0  &  1\\
   1  &  1  &  1  &  1  &  1 &   1   &      0
\end{bmatrix} \ \text{and} \ 
{\bs \omega} = 0.1\cdot \begin{bmatrix}1\\
   2\\
    3\\
    4\\
  5\\
   6\\
  7\end{bmatrix},
$$
where the entry highlighted in red is the one causing the asymmetry in the network coupling. The target phase differences $ x_{1i}$ are ${\bs x}_\mathrm{desired} = \left[\frac{\pi}{3} ~\frac{\pi}{4} ~\frac{\pi}{6} ~\frac{\pi}{8} ~\frac{\pi}{8} ~\frac{\pi}{6}\right]^\transpose$, and the desired phase locking frequency is $\omega_\mathrm{sync} = 1$ rad/sec.
The solution $\bs\alpha^*$ to problem~\eqref{eq: optimization 3} above yields
$$
A_\mathrm{d} = \begin{bmatrix}
0 & 0 & -1.2238 & 1 & 1 & 1 & 1 \\
    -3.7832  &       0 & 1  &  1  &  1  &  1  &  1\\
    -2.4384 &  1    &     0  &  1  &  1  &  1  &  1\\
  0.3978 &  1.6022  &   1   &      0  &  1  &  1 &   1\\
   1  &  0.3925  &  1  &  1     &    0  &  1  &  1\\
    1  &  0.2282  &  1  &  1  &  1    &     0  &  1\\
   0.6978  &  1.3022  &  1  &  1  &  1 &   1   &      0
\end{bmatrix}.
$$
Supplementary Fig.~\ref{fig: asymmetry} illustrates that the phase trajectories associated with such a solution achieve the target functional pattern.

We conclude this discussion by remarking that, besides our optimization problems, the sufficient conditions (i.a), (i.b), and (i.c) in the main text for the existence of positive coupling strengths that realize a target pattern can also be adapted to the case of directed networks. To show this, we let ${\bs\omega}_\mathrm{d} = [\omega_1-\omega_\text{sync}~\cdots~\omega_n-\omega_\text{sync}]^\transpose$ and $\bar B_\mathrm{source} = B_\mathrm{source}~\mathrm{sign}(D_\mathrm{d}(\bs x))$. Then, a sufficient condition for the existence of positive network weights that achieve a desired functional pattern is the following one.
\smallskip

\emph{There exists ${\bs \delta}\ge0$ such that $ B_\mathrm{source} D_\mathrm{d}({\bs x}){ \bs \delta_\mathrm{d}} =  { \bs  \omega}_\mathrm{d}$ if there exists a set $\mc S$ satisfying:}
\begin{enumerate}
\item[(iii.a)] \emph{$D_{\mathrm{d}_{ii}}(\bs x)D_{\mathrm{d}_{jj}}(\bs x)B_{\mathrm{source}_{:,i}}^\transpose B_{\mathrm{source}_{:,j}} \le 0$ for all $i,j\in \mc S$ with $i\ne j$;}
\item[(iii.b)] \emph{${ \bs  \omega}_\mathrm{d}^\transpose B_{\mathrm{source}_{:,i}} D_{\mathrm{d}_{ii}}(\bs x) > 0$ for all $i\in\mc S$;}
\item[(iii.c)] \emph{${\bs   \omega}_\mathrm{d} \in \mathrm{Im}( B_{\mathrm{source}_{:,\mc S}})$.}
\end{enumerate}

Note that, since $\omega_\mathrm{sync}$ is not known \emph{a priori} in most cases, these conditions are hard to check.
}

\subsection{Coupled Kuramoto oscillators to approximate fMRI data}
{
The interaction between static large-scale structural architecture of the human brain and local oscillations of neural communities is a key factor in the functional connectivity patterns that are empirically observed through functional magnetic resonance imaging (fMRI) when the brain is in a resting-state condition \cite{MPV-HEHP:10}. In the last two decades, extensive literature has resorted to Kuramoto phase oscillators to model fMRI data \cite{GD-VK-ARM-OS-RK:09,Corbetta2015,AP-MR:15,JC-EH-OS-GD:11,FV-MS-PJH-GS-JC-RL:15,PH-AV-PL-LM-VV:18,TM-GB-DSB-FP:19b}.
Many works, such as Ref.~\cite{Corbetta2015} and Ref.~\cite{JC-EH-OS-GD:11}, focus on the analysis of the oscillatory behaviors of neural populations that lead the emergence of functionally connected networks by modeling fMRI data as the output of networks of Kuramoto oscillators. The main working assumption is that at each node of the structural brain network there exists a community of excitatory and inhibitory neurons whose dynamical state is in a regime of self-sustained oscillations. From a modeling standpoint, this assumption is equivalent to employing a network of weakly coupled Wilson-Cowan oscillators \cite{FCH-EMI:97,AD-VWB:11}, or to a supercritical Andronov-Hopf bifurcation, such as the Stuart-Landau model in oscillatory regime \cite{MJY-LU:2015}. In this setting, the neurons' firing rates describe a closed periodic trajectory in phase space; that is, the firing rates delineate a limit cycle. Thus, the dynamics can be approximated by a single variable, which is the angle (or \emph{phase}) on this cycle. This regime is then modeled by a network of coupled heterogeneous Kuramoto oscillators that are connected to each other according to the architecture of the human brain.
}

\subsection{Procedure to extract phase-locked trajectories from fMRI data}

Following the procedure in Ref.~\cite{Corbetta2015}, we apply a narrow-band filter in the low frequency range $[0.04~0.07]$ Hz to the time series for each brain region. Next, to obtain a measure of the functional synchrony between the brain regions, we generate the $n\times n$ functional connectivity matrix ${F}$, whose entry ${F}_{ij}$ indicates the pairwise Pearson correlation coefficient between filtered time series of recorded neural activity.
To map these functional correlations to the phase domain, we extracted the phase time series $\tilde{\bs \theta}(t)$ by applying a Hilbert Transform to the filtered signals. From $\tilde{\bs \theta}(t)$, one can identify time windows over which frequency synchronization emerges with the aid of the phase-locking value matrix $P = [P_{ij}]$, where
$$
P_{ij}(t_0,t_f) = \frac{1}{t_f-t_0} \sum_{t=t_0}^{t_f} \left| e^{\mathsf{i}[\theta_j(t)-\theta_i(t)]} \right|.
$$
Clearly, if $P_{ij}(t_0,t_f) \approx 1$ for all $i,j$, then the time window $[t_0~t_f]$ comprises phase-locked trajectories. 

Since the phase time series $\tilde{\bs \theta}(t)$ are derived from inherently noisy measurements, we compute the best estimate of the phases ${\bs \theta}^*$ (modulo rotation) that are compatible with the noisy measurement in $\tilde{\bs \theta}(t)$ by solving the \emph{nonconvex phase synchronization} problem \cite{NB:16} -- that is, the estimation of phases from noisy pairwise relative phase measurements. Given a time window of frequency-synchronized phase time series $\tilde{\bs \theta}$, we find that $R \approx F$ (see main text and Fig.~\ref{fig: FC-R}).
Moreover, it holds that $\|R -  F\|_2 \to 0$ as $ P_{ij} \to 1$ element-wise. This implies that functional relationships between the phases $\bs \theta^*(t)$ (encoded in the matrix $R$) represent the same functional relationships that are measured in fMRI data (encoded in the matrix $F$), supporting the usage of Kuramoto oscillators to analyze neural synchronization.
  
 \subsection{Power network modeling and assumptions}
The main manuscript contains an application of our network tuning methods to the IEEE 39 New England power distribution network \cite{TA-RP-SV:79,PWS-MAP:98}. To model the dynamics of this network, we consider a connected power network with generators $\mc V_1$ and load buses $\mc V_2$.  A structure-preserving power network model contains $|\mc V_1|$ second-order Newtonian and $|\mc V_2|$ first-order kinematic phase oscillators obeying \cite{PWS-MAP:98}:
\begin{align}\label{eq: second order}
\begin{cases} M_i \ddot \theta_i + D_i \dot \theta_i & = \omega_i + \sum_{j=1}^{|\mc V_1|} a_{ij} \sin(\theta_j-\theta_i), \ i\in \mc V_1, \\[.3em]
D_i \dot \theta_i & = \omega_i + \sum_{j=1}^{|\mc V_2|}  a_{ij} \sin(\theta_j-\theta_i), \ i\in \mc V_2,
\end{cases}
\end{align}
where $M_i$, $D_i$ are the generator inertia constant, and the damping coefficient, respectively. In the equation for the generators $\mc V_1$, $ \omega_i = P_{m,i}$, which is the mechanical power input from the prime mover, and in the equation for the load buses $\mc V_2$, $\omega_i = P_{\ell,i}$, which denotes the real power drawn by load $i$. Finally, the weight $a_{ij}$ equals $a_{ij} = |v_i||v_j|\mathrm{Im}({Y}_{ij})$, with $v_i$ denoting the nodal voltage magnitude and ${Y}_{ij}$ being the admittance matrix. The above structure-preserving power network model represents an AC grid with a synchronous generator.

Owing to { Ref.~\cite[Lemma 1]{FD-MC-FB:13}}, the existence and local exponential stability of synchronized solutions of the oscillator model Eq.~\eqref{eq: second order} can be entirely described by means of the first-order Kuramoto model. That is, the load dynamics of a structure-preserving power grid model has the same stable synchronization manifold of Eq.~(1) in the main text.

We assume that thermal limit constraints are equivalent to phase cohesiveness requirements. To be precise, we obtain a bounded power flow $a_{ij}\sin(\theta_j-\theta_i)$ for the line $(i,j)$ whenever the angular distance $|\theta_j-\theta_i|$ is bounded, which is satisfied by frequency-synchronized phase trajectories.
Moreover, we assume constant voltage magnitudes $|v_i|$ at the loads, so that the weights $a_{ij}$ can be considered fixed. This is a standard assumption in power systems ({ also known as} \emph{decoupling assumption}). We refer the interested reader to { Ref.~\cite[Remark 1]{FD-MC-FB:13}} for further details.
 
The generators and bus parameters for the IEEE New England Power network are available in the original article and in classic textbooks \cite{TA-RP-SV:79,PWS-MAP:98}. { For simplicity, we set $D_i=1$ for all loads, which corresponds to a highly damped scenario, possibly due to local excitation controllers}. In our simulations, we utilized the standard optimal power flow solver provided by MATPOWER to compute the parameters $\bs v$, $\bs p_\ell$ and ${\bs \theta}(0)$ needed to integrate the Kuramoto model in Matlab through a standard \texttt{ode45} solver.
 
 \subsection{Application to additional power network models}
 
 {
 Depending on which assumptions are made and on which application is studied, there exist many different power network models in the literature. To demonstrate that the fundamental principles of our procedure remain unchanged even when dealing with more complex dynamics that relax some of the modeling assumptions, we apply our method to two models that differ from the one in equations~\eqref{eq: second order}. Henceforth, our goal is to intervene on a power network after a fault occurs between two loads in order to recover a desired (pre-fault) functional pattern.
 
First, we study the case of a third-order (also known as one-axis) model, such as the one in Ref.~\cite{KS-LRG-MM-TF-DW:18}. The main differences with the structure-preserving power network in equations~\eqref{eq: second order} is that the model in Ref.~\cite{KS-LRG-MM-TF-DW:18} includes the transient dynamics of voltage magnitudes, and that electrical loads are simply modeled as passive impedances. For $N=10$ generators, Supplementary Fig.~\ref{fig: kron reduced}{\bf a} illustrates the reduced power network model obeying the dynamics \cite{KS-LRG-MM-TF-DW:18}
\begin{align}\label{eq: third order}
\begin{cases}
\dot\theta_i = \omega_i,\\
M_i\dot\omega_i = p_{m,i} - D_i\omega_i +\sum_{j=1}^N |v_i| |v_j| \mathrm{Im}({Y}_{ij}) \sin(\theta_j-\theta_i),\\
T_i\dot v_i = v_i^{\mathrm{f}} - v_i +(\chi_i^\prime - \chi_i) \sum_{j=1}^N |v_j| \mathrm{Im}({Y}_{ij}) \cos(\theta_j-\theta_i),
\end{cases}
\end{align}
where $\theta_i$ is the rotor angle, $\omega_i$ its frequency, $p_{m,i}$ is the effective mechanical input power of the machine $i$, $M_i$ and $D_i$ are the inertia and damping of the mechanical motion, respectively, $v_i$ indicates the transient voltage, $\mathrm{Im}({Y}_{ij})$ is the susceptance of the transmission line $(i,j)$, $T_i$ denotes the relaxation time of the transient voltage dynamics along the $q$ axis, $v_i^{\mathrm{f}}$ is the internal voltage, and, finally, $\chi_i$ and $\chi_i^\prime$ are the static and transient reactances along the $d$-axis.

Akin to the structure-preserving power network model, stationary operation of the grid corresponds to constant voltages and frequencies, along with constant rotor phase differences. Since our procedure relies on phase-locked trajectories to achieve a desired functional pattern (i.e., power flow), it can be adjusted to intervene on the stationary operation of a challenging model such as the one in equations~\eqref{eq: third order}. In fact, in regimes with small voltage and frequency swings, the system is still well approximated by first-order Kuramoto dynamics. Clearly, for a desired functional pattern, the error between the one estimated from the application of our procedure and the one obtained from the dynamics in equations~\eqref{eq: third order} will be proportional to the changes in voltages and frequencies.

To test our approach on the model from equations~\eqref{eq: third order}, we use the same IEEE 39 New England benchmark case as in the main manuscript. The initial network parameters for our simulations, which represent standard grid operating conditions, are taken from Ref.~\cite{YS-IM-TH:11} and Ref.~\cite{AM-IK-PB-GS:15}. The voltage $v_i(0)$ and the initial condition for $\theta_i(0)$ and $\omega_i(0)=0$ for generator $i$ are fixed using power flow computation. The goal of this application is to recover the initial power flow after the same fault as in Ref.~\cite{YS-IM-TH:11} (a line trip between loads 16 and 17) occurs. Such a fault affects the network admittance matrix, and yields an undesired power flow.
To recover the pre-fault power flow we follow the same steps as described in the main text (see also Fig.~5{\bf b}), but because the reduced-network structure is typically a complete graph \cite{FD-FB:12b}, we do not modify its coupling strengths. Instead, we assume that we can adjust the values of $\bs p_{m}\in\real^{N}$, so that $\bs p_{m}-\mathrm{diag}(D_1, \dots, D_N)\bs \omega$ are the oscillators' natural frequencies that are tuned in order to achieve the pre-fault functional patterns in the post-fault network. The values for $P_{m}$ are computed by rewriting the second one of equations~\eqref{eq: third order} as equation~(4) in the main text, and solving for the natural frequencies. For the coupling values in $\bs \delta$, we use $a_{ij} = |v_i^\mathrm{ss}||v_j^\mathrm{ss}|\mathrm{Im}({Y}_{ij})$, where $v_i^\mathrm{ss}$ denotes the steady state value of the voltage $v_i$ before applying our intervention. Finally, we sum a positive constant to the obtained natural frequencies so that they are all positive -- this constant can be used to adjust the generators to a desired average mechanical input.

Supplementary Fig.~\ref{fig: kron reduced}{\bf b}-{\bf c} illustrate that our procedure is able to adequately recover the desired functional pattern after the line trips occurs. The error between the pre-fault pattern $R_0$ and the recovered one $R_\mathrm{recovered}$ is due to small changes in the frequency and voltage values after $P_m$ is modified. We remark that our procedure relies on the dynamics of $\bs \omega$ and $\bs v$ leading to small changes. In situations where these changes significantly affect the phases dynamics, the first-order Kuramoto approximation leveraged by our method may not successfully recover a desired functional pattern.
\medskip

We now turn our attention to models that do not neglect energy losses, such as the one in Ref.~\cite{FH-PS-RL-TK-JK-YM:20}. By compensating for the losses in the injected power but considering lossy interconnections, the structure-preserving model in equations~\eqref{eq: second order} becomes,
\begin{align}\label{eq: lossy}
\begin{cases} M_i \ddot \theta_i + D_i \dot \theta_i & = \omega_i + \sum_{j=1}^{|\mc V_1|} a_{ij} \sin(\theta_j-\theta_i+\varphi), \ i\in \mc V_1, \\[.3em]
D_i \dot \theta_i & = \omega_i + \sum_{j=1}^{|\mc V_2|}  a_{ij} \sin(\theta_j-\theta_i+\varphi), \ i\in \mc V_2,
\end{cases}
\end{align}
where $\varphi$ represents the phase shift induced by energy losses.
 We show in Fig.~\ref{fig: error lossy network} that our procedure to restore a desired functional pattern in a lossy power network after a fault still works well for small values of the phase shift $\varphi$. We can recover functional patterns associated to target active power flow conditions by modifying the constraints of our optimization procedures to explicitly take into account the dynamics in equations~\eqref{eq: lossy}.
 
 At synchronous operating conditions, the dynamics in equations~\eqref{eq: lossy} reduce to the ones of  Kuramoto-Sakaguchi oscillators \cite{HS-YK:86}:
$$
\dot \theta_i = \omega_i + \sum_{j=1}^n A_{ij}\sin(\theta_j - \theta_i + \varphi).
$$
The phase shift $\varphi$ makes the matrix formulation of the above dynamics to be incompatible with the ones we have introduced in equation~(3) of the main text. We can instead write the above dynamics in matrix form by considering the graph $\mc G$ as a directed graph (see also Supplementary Text~\ref{sec: asymmetry} above), with $\mc E_\mathrm{d}$ being the set of directed edges. That is, we consider each undirected edge as two directed edges where each direction $(j,i) \ne (i,j)$ has the same weight $A_{ij}=A_{ji}$.
Then, we can write the following equation:
\begin{equation}\label{eq: KS matrix}
[\omega_1~\cdots~\omega_n]^\transpose + B_\mathrm{sink} D_\mathrm{d} ({\bs x},\varphi) [\bs \delta^\transpose~\bs \delta^\transpose]^\transpose = \omega_\mathrm{sync}\1,
\end{equation}
where $B_\mathrm{sink}\in\real^{n\times|\mc E_\mathrm{d}|}$ satisfies $B_{\mathrm{sink},{k\ell}}=-1$ if $k$ is the sink of the interconnection $\ell$ and $B_{\mathrm{sink},{k\ell}}=0$ otherwise, $D_\mathrm{d}\in\real^{|\mc E_\mathrm{d}| \times \mc |\mc E_\mathrm{d}|}$ is the diagonal matrix of all $\sin( x_{ij}+\varphi)$, and $\omega_\mathrm{sync}\in\real$ is the synchronization frequency of phase-locked trajectories.

We are now ready to apply the equation~\eqref{eq: KS matrix} as a constraint in our numerical optimization routine. Without loss of generality, we set $\omega_\mathrm{sync} = 0$, and apply the same method developed for the lossless case to the IEEE 39 New England test case to compute the optimal correction after a fault occurs between loads 13 and 14 (the same as in the main text). For a loss $\varphi = 0.01$, the updated optimization is able to recover the pre-fault functional pattern with a mean error $<\mathrm{vec}(R) - \mathrm{vec}(R_\mathrm{recovered})> =0.072$, where $\mathrm{vec}(\cdot)$ vectorizes the matrix (see also Supplementary Fig.~\ref{fig: recovery lossy}). This result improves upon the original method proposed for lossless  networks by guaranteeing a satisfactory active power flow recovery for losses $\varphi$ that are one order of magnitude larger.
Finally, we observe that for $\varphi>0.01$ the fixed parameters of this specific system cause the phases to lose frequency synchronization even at operating conditions.
 }

%\bibliographystyle{plain}
%\bibliography{alias,FP,Main,New}
\vfill
 
 \newpage
 \section{Supplementary Figures}
 
  \vspace{4cm}
\vfill
\begin{figure}[h]
\centering
\includegraphics[width=.9\textwidth]{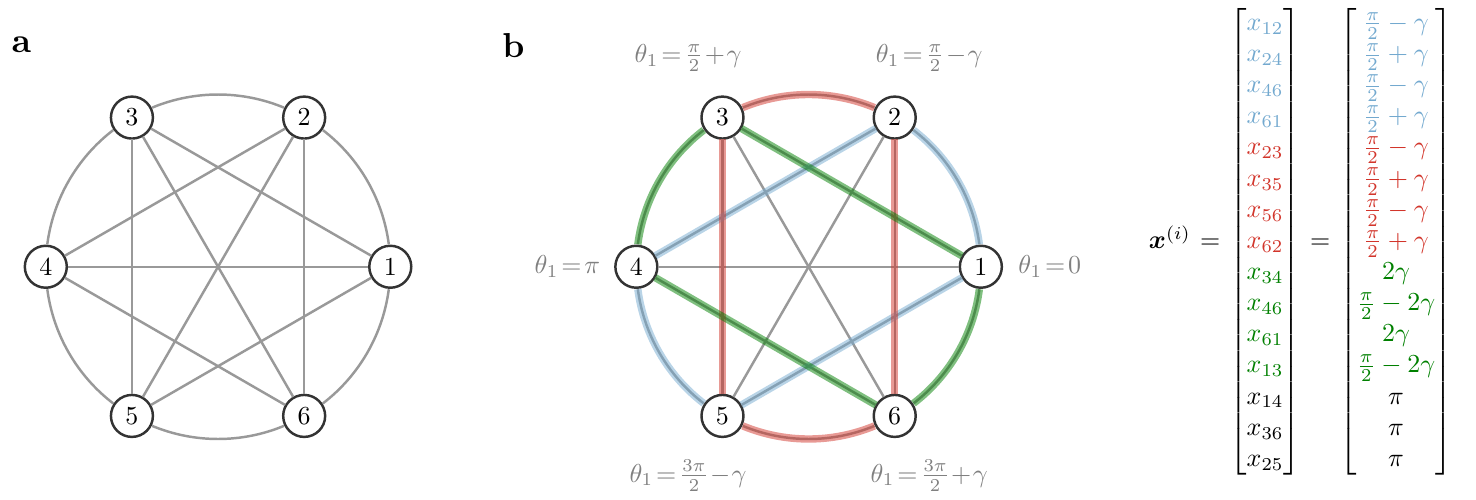}
 \vspace{0.4cm}
\caption{{ {\bf Infinite compatible patterns in a complete graph of identical oscillators.} {\bf a} A network described by a complete graph of $n=6$ oscillators with interconnection weights $\boldsymbol \delta = \bs 1$ and homogeneous natural frequencies $\bs\omega = \boldsymbol 0$. {\bf b} Each cycle of the network is highlighted in a different color, which is reflected on the entries of the phase differences equilibria $\bs x^{(i)}$. It holds $\sin(\bs x^{(i)}) = [a~a~a~a~a~a~a~a~b~b~b~b~0~0~0]^\transpose \in\mathrm{ker}(B)$, where $a = \sin(\frac{\pi}{2}-\gamma) = \sin(\frac{\pi}{2}+\gamma)$ and $b = \sin(2\gamma) = \sin(\pi-2\gamma)$ for all $\gamma\in(0,\frac{\pi}{2})$. Clearly, any value for $\gamma$ in the latter interval determines a distinct compatible functional pattern.} \label{fig: complete graph}}
\end{figure}
\vfill

\begin{figure}[h]
\vspace{3cm}
\centering
\includegraphics[width=.9\textwidth]{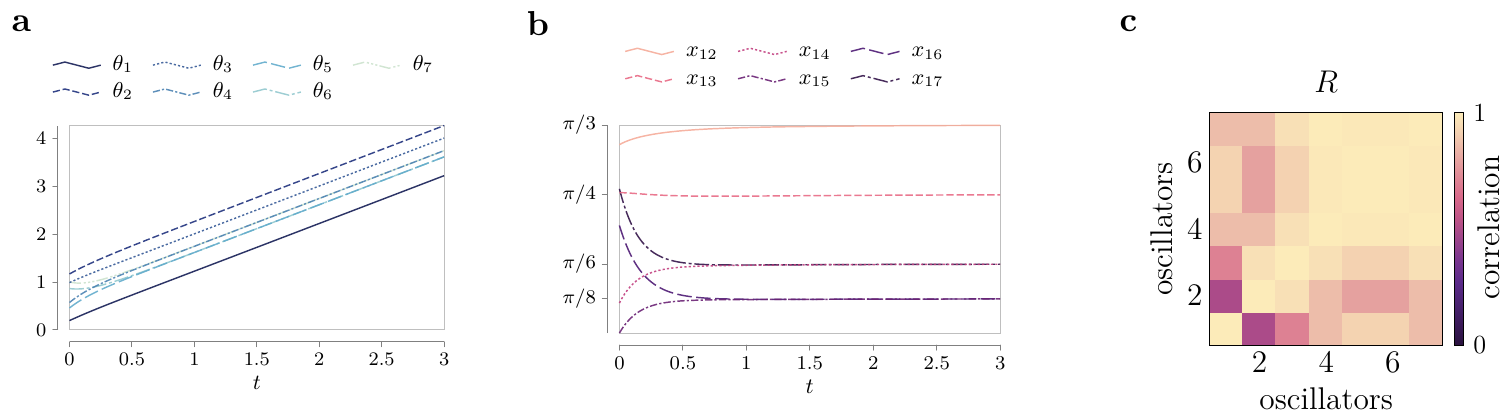}
 \vspace{0.4cm}
\caption{ {\bf Desired functional pattern and phase-locked frequency in a directed network.} {\bf a} Phase trajectory of the modified network after the solution $\bs\alpha^*$ to the problem in \eqref{eq: optimization 3} is applied to a network of $n=7$ oscillators with target phase differences $ x_{1i}$ equal to ${\bs x}_\mathrm{desired} = \left[\frac{\pi}{3} ~\frac{\pi}{4} ~\frac{\pi}{6} ~\frac{\pi}{8} ~\frac{\pi}{8} ~\frac{\pi}{6}\right]^\transpose$ and desired phase locking frequency is $\bar k_\mathrm{freq} = 1$. The initial conditions  $\bs x_0$ are chosen randomly and satisfy $\|\bs x_0 -{\bs x}_\mathrm{desired} \|<0.5$. {\bf b} The phase difference trajectories achieve the target values in ${\bs x}_\mathrm{desired}$. {\bf c} The phase trajectories achieve the functional pattern associated with the target phase differences. \label{fig: asymmetry}}
\end{figure}
\vspace{2cm}

\begin{figure}[htb]
\centering
\includegraphics[width=1\textwidth]{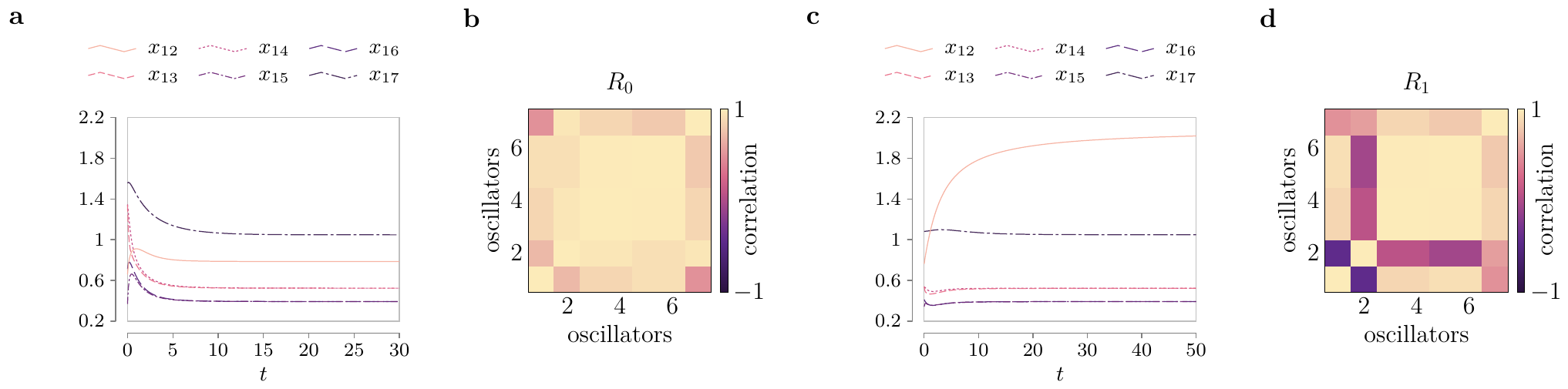}
\vspace{0.2cm}
\caption{{\bf Results of the heuristic method to promote stability of functional patterns containing negative correlations.} \textbf{a} Phase differences of the network with adjacency matrix $A$ in Eq.~\eqref{eq: initial}. The phase differences converge to $ {\bs x}_\text{desired}^{(0)}$. \textbf{b} The functional pattern associated with the phase differences in panel \textbf{a}. \textbf{c} Phase differences of the network with adjacency matrix $\tilde A$ in Eq.~\eqref{eq: adjusted}, after the optimal adjustment is computed from Eq.~\eqref{eq: heuristic}. The phase differences converge to $ {\bs x}_\text{desired}^{(1)}$. \textbf{d} The functional pattern associated with the phase differences in panel \textbf{a}. Notice that the only rows and columns that change from the functional pattern $R_0$ are the second row and second column. This is due to the fact that only $x_{12} = \theta_2-\theta_1$ differs between the two equilibria $ {\bs x}_\text{desired}^{(0)}$ and $ {\bs x}_\text{desired}^{(1)}$. \label{fig: heuristic}}
\end{figure}
\vspace{4cm}

\begin{figure}[t]
\centering
\includegraphics[width=1\textwidth]{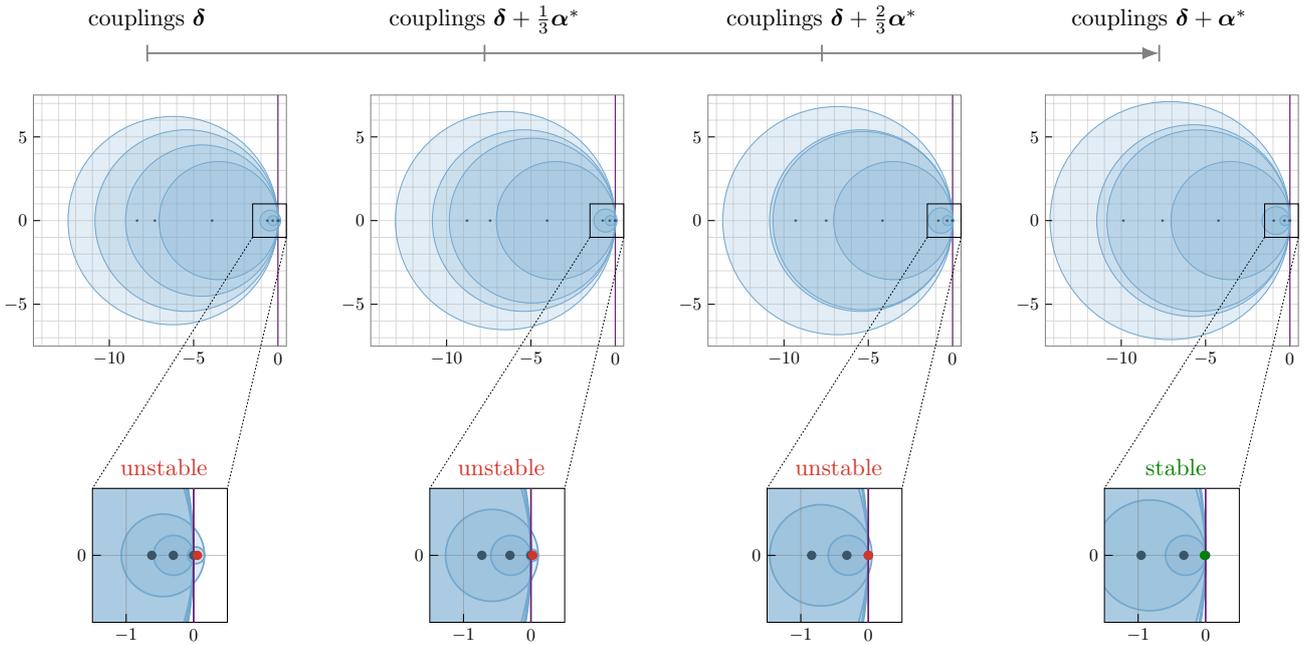}
 \vspace{0.4cm}
 \caption{ \textbf{Refined mechanism underlying the heuristic procedure to
     promote stability of functional patterns containing negative
     correlations.} For the $7$-oscillator network in Supplementary
   Text 1.5, we apply the procedure in equation~\eqref{eq: heuristic2}
   with $c_1=0.1$ and $c_2=10$ to achieve the stability of the pattern
   $ {\bs x}_\text{desired} = \left[\frac{21\pi}{32} ~ \frac{\pi}{6}
     ~\frac{\pi}{6} ~\frac{\pi}{8} ~\frac{\pi}{8}
     ~\frac{\pi}{3}\right]^\transpose$, where
   $x_{12} = \theta_2-\theta_1 >\frac{\pi}{2}$. Notice that, differently from Fig.~7 in the main text, the minimization in equation~\eqref{eq: heuristic2} enables a refined optimization of the network weights through the scaling parameters $c_1=0.1$ and $c_2=10$. The left plot
   illustrates the Gerschgorin disks and the Jacobian's eigenvalues
   locations for the original network. It can be observed in the
   zoomed-in panel that one eigenvalue is unstable
   ($\lambda_2 = 0.0565$, in red). The optimal correction
   $\bs\alpha^*$ of the oscillators' coupling strengths is gradually
   applied from the left-most panel to the right-most one at
   $\frac{1}{3}$ increments. The right zoomed-in panel shows that, as
   a result of our procedure, $n-1$ eigenvalues ultimately lie in the
   left-hand side of the complex plane ($\lambda_1=0$ due to
   rotational symmetry and $\lambda_2=-0.0178$, in green). \label{fig:
     eig shift2}}
\end{figure}
\vspace{3cm}

\begin{figure}[h]
\centering
\includegraphics[width=.7\textwidth]{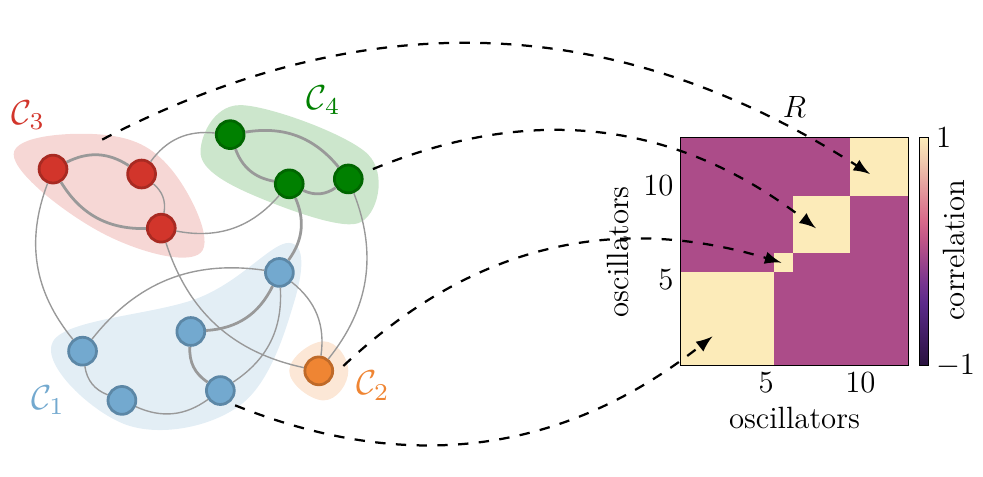}
 \vspace{0.5cm}
\caption{\small  {\bf A network of $n=12$ oscillators with partition $\mc C = \{\mc C_1, \dots, \mc C_4\}$ and the functional pattern $R$ associated with cluster-synchronized trajectories.} Each cluster consists of synchronized oscillators, which produce a correlated diagonal block in the pattern $R$ that satisfy $\rho_{ij}=1$ for all $i,j \in\mc C_k$, $k\in\{1,2,3,4\}$. \label{fig: cluster synch}}
\end{figure}
\vspace{3cm}

 \begin{figure}[htb]
  \vspace{4cm}
\centering
\includegraphics[width=.8\textwidth]{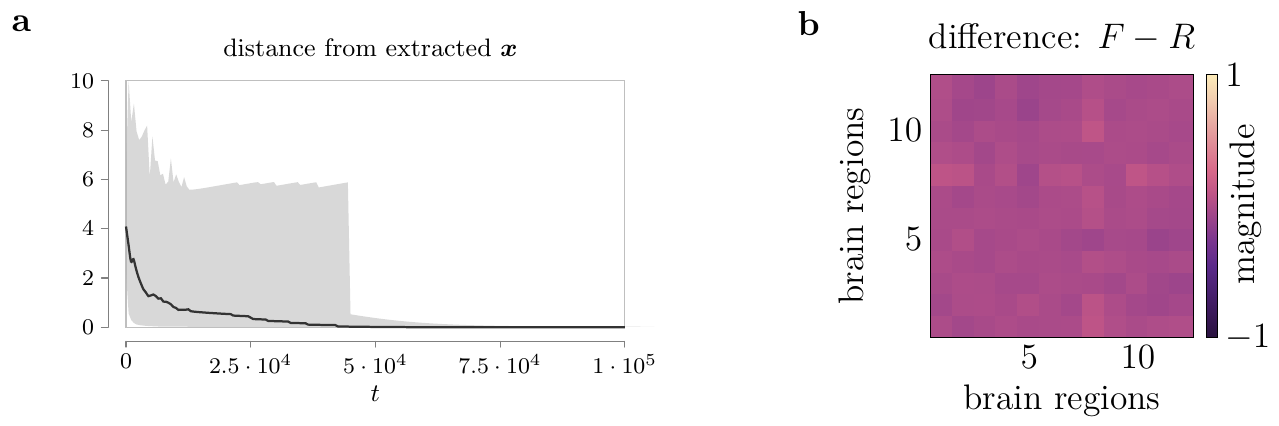}
 \vspace{0.3cm}
\caption{{\bf Additional analysis on the functional pattern $R$ in the brain network application.}  { \textbf{a} Stability of the prescribed functional pattern $R$ from random initial conditions $\bs x_0$ satisfying $\|\bs x_0- {\bs x}\|_\infty\le\frac{\pi}{2}$. The thick black line represents the average $\ell_2$-norm distance over $10^4$ random initializations, and the shaded area represents the smallest and largest value of the $\ell_2$-norm distance. The sudden drop in the largest norm is due to the phases $\bs \theta \in\mathbb{T}^n$ evolving in the torus, thus taking values in the interval $[0~2\pi)$. Our numerical simulations reveal that phase trajectories starting from the other half of the torus (i.e., $\|\bs x_0- {\bs x}\|_\infty>\frac{\pi}{2}$) may converge to different stable patterns, which are not compatible with the phases extracted from the functional MRI recordings.} \textbf{b} The difference between the functional connectivity $F$ and the functional pattern $R$. The entries with the largest magnitude are $\approx 0.08$, highlighting the stark similarity between the two correlation patterns $R$ and $F$. \label{fig: FC-R} }
\end{figure}

\begin{figure}[t]
\centering
\includegraphics[width=1\textwidth]{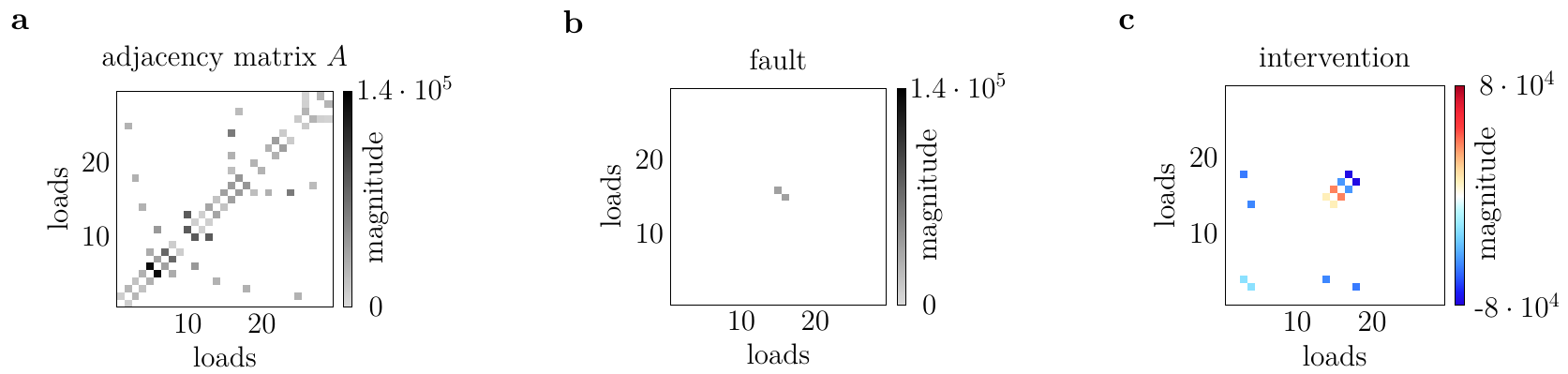}
 \vspace{0.2cm}
\caption{{\bf Matrices that describe the network interconnections, the fault, and the local intervention of the power network parameters to recover the pre-fault power distribution.} \textbf{a} The adjacency matrix used in the Kuramoto model to simulate the IEEE 39 power network. \textbf{b} The fault that disconnects loads $13$ and $14$. \textbf{c} The intervention is localized, in the sense that only branches of the loads connected to the ones affected by the fault and their immediate neighbors require adjustments. The sparsity of the local intervention is promoted by the usage of the $\ell_1$-norm in the optimization problem. \label{fig: modification}}
\end{figure}
\vspace{2cm}

\begin{figure}[h]
\centering
\includegraphics[width=1\textwidth]{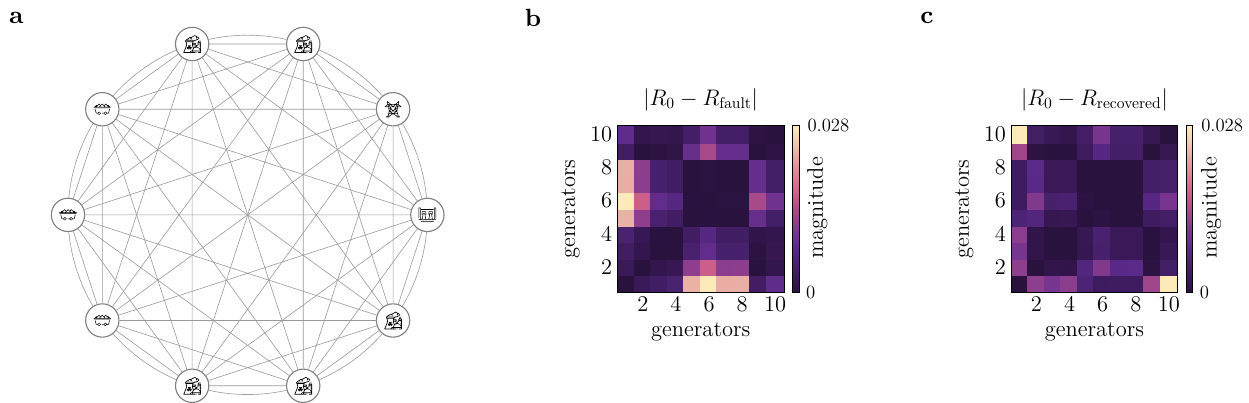}
 \vspace{0.2cm}
\caption{ {\bf Application of our procedure to the third-order model of the IEEE 39 New England.} {\bf a} The IEEE-39 New England power network reduced to a 10-generator network. Electrical loads are simply modeled as passive impedances. In order to explicitly account for the outside of the system, Generator 1 is assumed to be connected to an infinite bus and has constant phase and frequency \cite{YS-IM-TH:11}. {\bf b} The absolute difference between the pre-fault functional pattern $R_0$ and the post-fault pattern $R_\mathrm{fault}$. The Frobenius norm of this difference is $\|R_0-R_\mathrm{fault}\|_\mathrm{F} = 0.0907$. {\bf c} The absolute difference between the pre-fault functional pattern $R_0$ and the recovered pattern $R_\mathrm{recovered}$ after tuning the power $\bs p_m$ at the generators. The Frobenius norm of this difference is $\|R_0-R_\mathrm{recovered}\|_\mathrm{F} = 0.0653$. \label{fig: kron reduced}}
\end{figure}
\vspace{2cm}

\begin{figure}[h]
\centering
\includegraphics[width=.35\textwidth]{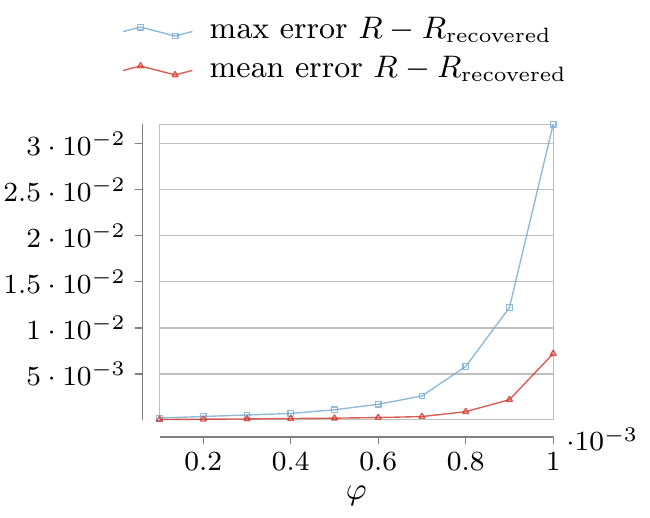}
 \vspace{0.4cm}
\caption{  {\bf Error between the pre-fault functional pattern $R$ and the functional pattern $R_\mathrm{recovered}$ obtained through our procedure as a function of the phase shift $\varphi\in\mathbb{S}^1$.} The functional pattern $R_\mathrm{recovered}$ is computed after a network correction due to a fault that occurs in the IEEE 39 test case between loads $13$ and $14$ (the same as in the main text). In the presence of a phase shift $\varphi$, the error between the desired functional pattern and the one associated with the network correction computed by our method remains small for small values of energy loss $\varphi$. Here, the parameter optimization does not explicitly account for the phase shift $\varphi$. The mean error is computed as $<\mathrm{vec}(R) - \mathrm{vec}(R_\mathrm{recovered})>$, and the maximum error as $\|\mathrm{vec}(R) - \mathrm{vec}(R_\mathrm{recovered})\|_\infty$, where $\mathrm{vec}(\cdot)$ denotes the vectorization. \label{fig: error lossy network}}
\end{figure}
\vspace{2cm}

\begin{figure}[h]
\centering
\includegraphics[width=.8\textwidth]{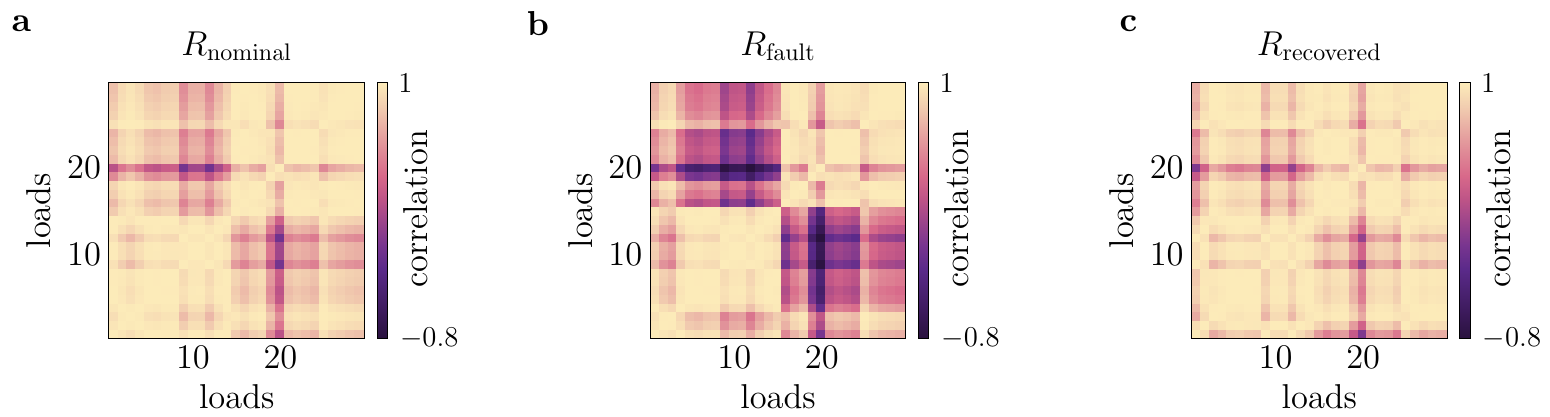}
 \vspace{0.4cm}
\caption{ {\bf Nominal, post-fault, and recovered functional pattern in a network with lossy communications.} In all panels, the loss is fixed to $\varphi = 0.01$. {\bf a} Functional pattern associated to nominal power flow in the IEEE 39 New England test case. {\bf b} Functional pattern associated to a power flow disruption due to a fault that disconnects loads 13 and 14. {\bf c} The recovered functional pattern after our procedure with updated constraint (equation~\eqref{eq: KS matrix}) is applied. The mean error between the pre-fault and the recovered functional patterns is $<\mathrm{vec}(R) - \mathrm{vec}(R_\mathrm{recovered})> = 0.072$, where $\mathrm{vec}(\cdot)$ denotes the vectorization of the patterns. \label{fig: recovery lossy}}
\end{figure}

\end{document}